\documentclass[preprint,aps,prc,nofootinbib,floatfix,showpacs]{revtex4}
\usepackage{epsfig,float}
\begin{document}

\title{Inclusive Pion Double Charge Exchange in
$^4$He at Intermediate Energies} 

\author{E. R. Kinney}
\altaffiliation[Present Address: ]{Department of 
Physics, University of Colorado, Boulder, CO 80309} 
\author{J. L. Matthews} 
\affiliation{Department of Physics and Laboratory for Nuclear Science \\
Massachusetts Institute of Technology, Cambridge, Massachusetts 02139}

\author{P. A. M. Gram}
\altaffiliation[Present Address: ]{82-1021 Kaimalu Place, Captain Cook, 
HI 96704}
\author{D. W. MacArthur}
\author{E. Piasetzky}
\altaffiliation[Present Address: ]{School of Physics and Astronomy, Tel 
Aviv University, Tel Aviv 69978, Israel}
\affiliation{Los Alamos National Laboratory, Los Alamos, New Mexico 87545}

\author{G. A. Rebka, Jr.}
\author{D. A. Roberts}
\altaffiliation[Present Address: ] 
{Radiation Oncology Department, University of Michigan, Ann 
Arbor, MI 48109}
\affiliation{Department of Physics, University of Wyoming, Laramie, 
Wyoming 82071}

\date{\today}

\begin{abstract} A systematic experimental study of inclusive pion double
charge exchange in $^4$He has been undertaken.  The reaction
$^4$He$(\pi^+,\pi^-)4p$ was observed at incident energies 120, 150, 180,
240 and 270 MeV; the $^4$He$(\pi^-,\pi^+)4n$ reaction was observed at
incident energies 180 and 240 MeV.  At each incident energy, the doubly
differential cross section was measured at three to five outgoing 
pion laboratory angles between
25$^\circ$ and 130$^\circ$. At each angle, cross sections were measured over 
the range of outgoing
pion energies from 10 MeV up to the kinematic limit for the
reaction in which the final state consists of the oppositely charged 
pion plus four free nucleons.

The spectra of outgoing pions are strikingly different from those observed
for the inclusive double charge exchange reaction in heavier nuclei, but 
resemble
those observed in the $(\pi^-,\pi^+)$ reaction in $^3$He.  The
forward-angle spectra in the $^3$He and $^4$He reactions exhibit a 
prominent peak at
high outgoing pion energies.  Interpretation of the peaks in $^3$He
($^4$He) as a three- (four-)nucleon resonance is ruled out by kinematic
analysis.  The results of a calculation, wherein the double charge
exchange reaction is assumed to proceed as two sequential single charge
exchange interactions, suggest that the high-energy peak is naturally
explained by this double scattering mechanism. Non-static treatment of the
$\pi N$ interactions and the inclusion of nuclear binding effects appear
to be important in reproducing the shape of the energy spectra at forward
angles.
 \end{abstract}


\maketitle

\section{INTRODUCTION}

Pion double charge exchange (DCX), a nuclear reaction in which two units of
charge are exchanged between a pion and a nucleus, is an important tool
with which to investigate the details of the pion-nucleus interactions as
well as correlations between nucleons within the nucleus.  DCX is unique
among pion-nuclear scattering reactions, as it must involve at least
two nucleons in order to conserve charge.  Thus, the physiognomy of the
DCX process revealed by the experimental cross section depends on the
coordinates of two particles in the wavefunction of the nucleus.  As a
probe of the pion-nucleus interaction, DCX may be used to isolate
contributions from different types of reaction mechanisms.  Other
pion-nucleus reactions appear to be dominated by single scattering
mechanisms, but DCX must proceed through less simple mechanisms such as
double scattering, as shown in Fig. 1(a).  

An experimental investigation of DCX in $^4$He at intermediate energies
(120--270 MeV) has been performed in order to study the multiple
scattering mechanisms of the $\pi$-nucleus reaction directly.  The choice
of $^4$He as a nuclear target allows an investigation of DCX in an
extensively studied few-body system for which microscopic treatment of the
dynamics is possible in principle.  At the same time, the $^4$He nucleus
has a large density which may allow some effects of the nuclear medium to
be investigated.  The final state is also relatively simple, consisting of
a pion and either four protons or four neutrons, resulting from,
respectively, the reactions $^4$He$(\pi^+,\pi^-)4p$ and
$^4$He$(\pi^-,\pi^+)4n$.

\section{Previous Experimental Work}

The first investigation of DCX in $^4$He was reported by Davis {\it et
al.}\cite{davis} in 1964.  This and several subsequent measurements
\cite{gilly,kauf,ungar} of DCX cross sections in $^3$He and $^4$He were undertaken
as searches for bound three- or four-neutron states using the $(\pi^-,\pi^+)$
reaction.  More broadly motivated studies of DCX in $^4$He at energies above the
$\Delta$ resonance were carried out using liquid helium bubble chambers
\cite{car,gail,jean}. In the energy range 98--156 MeV total cross 
section measurements were performed
by Falomkin {\it et al.}\cite{fal1,fal2} using a
high-pressure helium streamer chamber.  Although some interesting phenomena were
observed in these experiments, the limited statistics precluded quantitative
conclusions.  Ref. \cite{kth} contains a critical review of these and other
experiments on $^4$He.  Other early DCX experiments on various nuclei which used
nuclear emulsions, bubble chambers, and spark chambers as detectors are described
in Refs.\cite{wood,yuly}.

The first modern measurement of doubly differential cross sections for
inclusive DCX, using a magnetic spectrometer to detect the outgoing pions,
was made at the Schweizerisches Institut f{\" u}r Nuklearforschung (now
PSI) for the $^{16}$O$(\pi^+, \, \pi^- )$ reaction \cite{mischke}. 
Subsequently, the $^4$He$(\pi^+, \pi^-)$ reaction was studied by Stetz
{\it et al.}\cite{stetz1,stetz2} using the EPICS spectrometer at the Los Alamos
Meson Physics Facility (LAMPF).  
Stetz {\it et al.} measured doubly differential cross sections for three
incident
pion energies (140, 200, and 295 MeV) at several angles in the range
30$^\circ$--120$^\circ$.  One spectrum was a re-measurement of the
spectrum reported by Kaufman {\it et al.}\cite{kauf} at 140 MeV, and
confirmed that 
the normalization of the older data was incorrect (too small) by a
factor of 100, as had been suspected. Six additional
spectra were measured with pions at the two higher incident energies,
and one spectrum for the $^4$He$(\pi^-, \pi^+)$ reaction was observed at
295 MeV incident energy.

Total and differential cross sections and pion momentum distributions 
have been measured at six energies in the 70--130 MeV range using the 
CHAOS 
detector at TRIUMF\cite{grater}.  Some evidence for a $d'$ dibaryon 
state of mass 2.06 GeV is claimed to be found in the energy dependence 
of the total DCX cross section.

An extensive study of the $^3$He$(\pi^-, \pi^+)$ reaction\cite{yuly} in
the incident energy range 120 -- 240 MeV was carried out with the same
apparatus as that used in the experiment reported in this paper.  The
$^3$He data, with good statistical accuracy and complete
coverage of the outgoing pion energy spectrum, revealed a double-peaked
structure at forward angles \cite{yuly}.  This result is
in striking contrast to the spectra seen in a systematic study of DCX on
heavier nuclei ($^{16}$O to $ ^{208}$Pb)~\cite{wood}, in which this
structure is absent, and which roughly resemble the distribution of events
in four-body phase space.  A hint of this structure had been seen in the
earlier $^3$He DCX data of Sperinde {\it et al.}\cite {sper}, whereas it
was not apparent in the $^4$He data of Kaufman {\it et al.}\cite{kauf} or
those of Stetz {\it et al.}\cite{stetz2}.

\section{PREVIOUS THEORETICAL WORK}

The first microscopic calculation of the DCX process in $^4$He was
reported by Becker and Schmit\cite{bs}, using a reaction model developed
by Becker and Mari\'{c}\cite{bm}.  The sequential single-charge-exchange 
(SSCX) mechanism is
calculated in a fixed-nucleon (FN) framework, and only resonant
[$\Delta(1232)$] amplitudes are used to describe the $\pi N$ interaction,
with an off-shell correction factor.  Total
cross section predictions of this calculation are presented in Ref.
\cite{bb} and they are generally too large by an order of magnitude,
though
the energy dependence is reminiscent of that of the data, as will be
shown later.  The quantitative disagreement should be expected at the
very least from the use of plane waves to describe the outgoing pions,
i.e., the effects of absorption and multiple scattering are not included.

Germond and Wilkin\cite{gw,gw2} proposed that a major contribution
to DCX might come from a pion-pion scattering mechanism, as shown in Fig. 
1(b). The amplitude is
calculated for a nucleon pair, drawing on the current algebra theory of
Weinberg\cite{wein} for the $\pi \pi$ interaction.  The pion ``cloud"
between the nucleon pair is calculated with a relativistic expression
derived from the pseudoscalar pion and Dirac nucleon theory described by
Bjorken and Drell\cite{bjd}.  The relatively large total cross sections
obtained fueled speculation that this mechanism was important.  However,
the
mechanism shown in Fig. 1(c) must also be included, and it was found by
Robilotta and Wilkin\cite{rw} that its amplitude cancels that of the
mechanism in Fig. 1(b) to a large extent.  

Jeanneret {\it et al.}\cite{jean} considered a model of DCX
in which pion-induced pion production (PIPP) is followed by two-nucleon
absorption of one of the two final pions.  The calculation is a
convolution of the free PIPP cross section and pion absorption cross
section in deuterium $(\pi^+ d \rightarrow p p)$, with some corrections
for the available phase space.  At the high energies of their experiment
(1.5--2.0 GeV), the authors find general agreement of their predictions
with their measured total cross sections.

The calculation of Gibbs, Gibson, Hess, and Stephenson \cite{gibbs} was 
based on the same
reaction model as that of Becker and Schmit\cite{bs}, but brought to 
bear the more
sophisticated techniques pioneered by Gibbs and his
co-workers \cite{getal}.  It should be noted that a simpler 
approximation
of those methods was used in the DCX calculation; specifically, the 
coupled
scattering equations were not solved.  The initial and final pion 
states as
well as the final nucleon states were described with plane waves, 
although
the effect of simple FSI between the spectator nucleon pair was
investigated.  A pion scattering amplitude from fixed nucleon positions 
was
calculated, and then averaged by the $^4$He wavefunction.  The final
nucleon momenta were integrated over the available phase space using a
Monte Carlo technique.  

The total cross section predicted by this calculation is a factor of 
100 smaller than that measured in the energy region 98--156 MeV by 
Falomkin {\it et al.}\cite{fal1,fal2}.  This theoretical result is in fact 
consistent 
with the approximate agreement found with the Kaufman {\it et al.} 
measurement\cite{kauf} at 140 MeV, which is now known to have been 
incorrectly normalized by a factor of 
100.  The calculation also underpredicts the $0^{\circ}$ excitation 
function for 176 MeV pions measured by Gilly {\it 
et al.}\cite{gilly} and 
the pion 
momentum spectrum measured at 485 MeV by Carayannopoulos {\it et 
al.}\cite{car}; however, the authors \cite{gibbs} comment that 
the calculation is not expected to work well at 
the higher incident energies.

A series of papers containing theoretical predictions for DCX cross
sections in $^3$He and $^4$He was published by Jibuti and
Kezerashvili\cite{jk1,jk2,jk3,jk4,jk5,jk6}.  Their basic approach was 
to
expand initial and final nuclear wavefunctions in a hyperspherical
basis, solving four-body equations for the $^4$He nucleus using various
models of the $NN$ potential.  In Refs. \cite{jk1,jk2} these
wavefunctions were used to calculate DCX cross sections in $^4$He using
the exact same form of the double scattering operator as used by Gibbs
{\it et al.}\cite{gibbs}.  Surprisingly, agreement was found with both
the incorrect data of Kaufman\cite{kauf} and the presumably correct
data of Gilly {\it et al.}\cite{gilly}.  A similar calculation for DCX
in $^3$He was reported\cite{jk3} several years later, and successfully
reproduced the spectrum measured by Sperinde {\it et al.}\cite{sper},
explaining the high energy peak evident in their data as a result of
FSI.  The following year Jibuti and Kezerashvili published a long
report\cite{jk4} describing their wavefunction formalism and
calculation in some detail, drawing together their earlier work on DCX
in $^3$He and $^4$He, and extending the results by including more
hyperspherical basis states in their expansion of the nuclear
wavefunction.  In addition, the operator for the $\pi \pi$ scattering
mechanism was extracted without change from Germond and Wilkin\cite{gw}
and inserted between their wavefunctions.  Two predictions of DCX
energy spectra -- one using the $\pi \pi$ mechanism of Germond and
Wilkin and one using the SSCX mechanism of Gibbs -- were compared to 
one
of the spectra measured by Stetz {\it et al.}\cite{stetz2}.  Both
predictions agreed quite well with the data, and no combination of the
two was discussed, in this article or in later work\cite{jk5,jk6}.

A completely different approach to calculating inclusive reaction cross
sections has been taken by several other authors\cite{ht,oset}.  Cascade
calculations were performed which treated the propagation of
the pion through the nucleus classically, and used interaction
probabilities dependent on the local nuclear density.  H\" ufner and
Thies\cite{ht} derived an approximate solution of the Boltzmann 
transport
equation, and presented results for a number of different inclusive
reactions including the total cross sections for DCX in $^4$He.  The
resonant part of the $\pi N$ scattering amplitude was calculated
non-statically, using a Fermi gas distribution to describe the nuclear
density in coordinate and momentum space.  Binding effects were not
included, but the resonance width was modified to account for Pauli
blocking and absorption.  The results of their calculation agreed
qualitatively with the existing data, though they exceeded the 
measurement 
of Stetz {\it et al.}\cite{stetz1} and did not reproduce its energy
dependence.

Oset and co-workers\cite{oset} have emphasized the description of the
$\Delta$-nucleus interaction with density dependent terms calculated in
nuclear matter.  Their results for DCX in $^{16}$O, $^{40}$Ca,
$^{103}$Rh, and $^{208}$Pb were compared with data in Ref. \cite{wood}.
This calculation has been extended to nuclei as light as
$^6$Li\cite{fong}, although treating this system as ``nuclear matter" is
surely unrealistic.  No calculations are available for $^4$He.

More recently, the Gibbs {\it et al.}~calculation\cite{gibbs} was 
reviewed critically and modified by 
Rebka\cite{rebka} 
and Kulkarni\cite{kul}.  The phase space was calculated 
relativistically, an antisymmetrized initial wave function 
was used, the spin and isospin coefficients were recalculated, 
and the newer $\pi$-nucleon phase shift parametrization of Rowe, Salomon, 
and Landau\cite{rowe} was used.  This calculation yielded cross 
sections substantially larger 
than those of Ref. \cite{gibbs}, but still failed to reproduce 
either the magnitude or energy dependence of the Kaufman\cite{kauf} or 
Gilly\cite{gilly} cross 
sections. Calculations were also performed 
for the doubly differential cross sections measured 
in the present experiment, and the results will be compared with the data
in Section VII.  

Kulkarni\cite{kul} also constructed a relativistic three-body model of
$\pi$-nuclear interactions, and applied it to DCX.  This calculation 
will be described briefly and its results
compared with the present data in Section VII.

A recent paper by Alqadi and Gibbs \cite{alqadi} reported calculations 
of $^4$He($\pi^+, \pi^-$) cross sections using two different 
methods:  the two-nucleon
sequential single charge exchange model and an intranuclear cascade 
code.  Final state interactions were included, as well as the 
contribution of the 
pion-induced pion production process to the inclusive DCX cross 
sections at 240 and 270 MeV.  These calculations will be discussed 
in Section VII.

 \section{EXPERIMENTAL APPARATUS AND PROCEDURE}

In the present experiment, the inclusive DCX
processes $^4$He$(\pi^+, \pi^-)$ and $^4$He$(\pi^-, \pi^+)$ have been
investigated by measuring doubly differential cross sections, $d^2 \sigma
/ d\Omega dE$, at three to five outgoing pion laboratory angles in the 
range
25$^\circ$--130$^\circ$, for incident positive pions of energy 120, 150,
180, 240, and 270 MeV, and for incident negative pions of energy 180 and
240 MeV.  At each angle, measurements were performed for outgoing pion kinetic 
energies $E$ from 10 MeV up to the kinematic limit for the reaction.
Preliminary results of this experiment have been reported in Ref.
\cite{kprl}.

The experiment was performed with incident $\pi^\pm$ beams from the
high-energy pion channel (``P$^3$'') at LAMPF.  The experimental apparatus
and procedure have been discussed in detail by Wood {\em et al.}\
\cite{wood} and Yuly {\it et al.} \cite{yuly} and will be treated only
briefly here. The outgoing pions were detected using a 180$^\circ$
vertical bend, double focusing magnetic spectrometer, with an effective
solid angle of about 16 msr and a momentum bite of about 8\% (see 
Fig.~2).  The
180$^\circ$ bend of the magnet provides stringent selection of the charge
of the detected particle, and the short flight path (3.5 m), which was
entirely in vacuum except for a 2.5 cm interruption for a mid-
spectrometer wire chamber, allowed low-energy pions to reach the focal
plane with minimal corrections for scattering and decay.  Since no
detectors are required at the entrance to the spectrometer, high
luminosities could be used.

The detector system had five components: a wire chamber (WC0) placed in the
mid-plane of the spectrometer, two wire chambers (WC1 and WC2) near the
focal plane that were used to reconstruct trajectories, behind these, a 1.6
mm scintillator (S1), that provided an accurate time reference for the
trigger as well as pulse height information useful in the identification of
particles, and lastly a fluorocarbon (FC-88) \v Cerenkov detector that
distinguished electrons from pions. 

The trigger was a fourfold coincidence among the three wire chambers and
the scintillator.  Inclusion of the mid-spectrometer wire chamber
guaranteed that a particle had passed through the spectrometer, which
accomplished a crucial reduction of the trigger rate due to room
background. The timing information available from this chamber was also
used to distinguish pions from very slow protons. 

The liquid $^4$He target was a cylindrical cell 25 mm in diameter
and 75 mm high with a 51 $\mu$m thick Mylar wall.  It was surrounded by
four layers of 0.64 $\mu$m aluminized mylar super-insulation and, at a
radius of 19 cm, a 13 $\mu$m aluminum heat shield maintained at liquid
nitrogen temperature. The insulating vacuum of the cryostat was contiguous
with that of the spectrometer vacuum chamber. The top of the cell was
connected to the bottom of a 30 liter reservoir, which fed liquid $^4$He
to the target under the force of gravity.

The target cell and a CH (styrofoam) cylinder of the same dimensions were
mounted vertically on the axis of rotation of the spectrometer. An
elevator mechanism could position each of the targets in the path of the
beam, permitting convenient measurement of $\pi p$ scattering, which was
the reference cross section in this experiment.  Background from the
helium target cell walls was measured by detecting pions from the same
cell after the helium had been evacuated.  The measured yield from the
empty cell was typically less than 10\% of the yield measured when the
cell contained liquid helium. 

The temperature of the $^4$He was determined by the vapor pressure in the
reservoir, which was essentially the same as the local atmospheric
pressure (584 $\pm$ 13 mmHg) plus 0.5 lb/in$^2$ overpressure from a gas
outflow meter.  Interpolating from empirical measurements of liquid $^4$He
density as a function of vapor pressure one finds a density of $128.6 \pm
0.3$ mg/cm$^3$.  The $^4$He target density determination was checked in a
subsequent experiment on DCX in $^3$He \cite{yuly}.  Since at temperatures
below that necessary to liquify
$^3$He (2.17K) liquid $^4$He
becomes superfluid and no longer boils, its density is known more reliably.  The
density measurements reported in this paper were found to be in excellent
agreement with those made subsequently with the superfluid $^4$He target.

Before reaching the target, the pion beam passed through an
ionization chamber, which was
used to determine the relative flux. Since this device was sensitive to all
charged particles in the beam, it was necessary to normalize the flux
measurement each time the beam transport elements were adjusted. The size
(about 1.6 cm in diameter) and position of the beam spot were continuously
monitored by a small multi-wire proportional chamber placed downstream of
the ionization chamber and 30 cm upstream of the target. Downstream of the
$^4$He target, the pion beam struck a 6 mm thick polyethylene (CH$_2$) target.
Three-element plastic scintillator telescopes were placed on each side of
the CH$_2$ target to detect pions scattered at $90^\circ$. These telescopes
were used to monitor the position of the incident beam, as well as to
provide a check on the flux determined by the ionization chamber. The
response of the telescopes was sensitive essentially only to pions, since
beam contaminants are very unlikely to scatter at 90$^\circ$ into the
telescopes.

To measure the doubly differential cross sections the target was exposed to
an incident pion beam and events were collected at given spectrometer
momentum settings until statistical uncertainties of approximately 5\% in
the number of detected pions were achieved. Data were collected at 10 MeV
intervals in outgoing pion energy. In addition to each complete set of
$^4$He observations, a series of observations was made measuring the
background from the empty target walls. This process was repeated for each
spectrometer angle and incident energy. Each time the incident energy, and
hence the normalization of the beam monitors, was changed, a series of CH
normalizations, that covered the same range of angles but in smaller steps,
was performed.

\section{DATA ANALYSIS}

The goal of this experiment was to measure a doubly differential cross
section for each reaction, incident pion energy, scattering angle, and
outgoing pion energy studied. The doubly differential cross section is
related to observable quantities as follows:
\begin{equation}  \label{eqn:xsect}
\frac{d^2 \sigma}{d \Omega \, d E}= \frac{N_{{\rm det}} \, \epsilon_c}{
N_{{\rm inc}} \, \Delta \Omega \, \Delta E \, x \, \rho \, f_d \, f_l}
\end{equation}
where $N_{{\rm det}}$ is the number of pions detected, $N_{{\rm inc}}$ is
the number of pions that were incident, $x$ is the effective thickness of
the target, $\rho$ is the density of the target, $\Delta \Omega$ is the
effective solid angular acceptance of the spectrometer, $E$ is the outgoing 
pion 
kinetic energy and $\Delta E$ is its range, $\epsilon_c$ is a correction of the spectrometer acceptance
due
to multiple scattering and energy loss, $f_d$ is the correction due to pion
decay, and $f_l$ is the dead-time correction.

The analysis procedure is described in detail in \cite{kth} and will
only be summarized here. (See also ref\cite{yuly}.) 

\subsection{Wire chamber calibration and phase space definition}

Calibration constants relating time differences of signals from wire
chambers to position were established by placing a collimated $^{55}$Fe
source at precisely measured positions in front of the chambers. For each
event, the position information from WC1 and WC2 was used to reconstruct
the
particle trajectory back to the focal plane of the spectrometer.
Reconstructed trajectories were tested for conformity with the distribution
in phase space of particles that could have been transmitted by the 
spectrometer from the target to the focal plane.

\subsection{Particle Identification}

To calculate cross sections it was necessary to separate the pions from the
other particle species that also caused triggers. Protons were generally
eliminated by their large pulse heights in the scintillator. At a
spectrometer momentum setting of about 156 MeV/c, where the protons were 
just
reaching the scintillator, however, many of the protons deposited the same
amount of energy as the pions in the scintillator. At this momentum, the
protons
were distinguished by their longer time-of-flight between WC0 and S1.

Since pions, as well as electrons,\footnote{
Electrons or positrons, depending on the charge setting of the
spectrometer.
The word ``electron'' will be used generically throughout this discussion.}
emit \v {C}erenkov radiation at spectrometer settings greater than 180
MeV/c
($T_\pi =88$ MeV), it was not possible to use the \v {C}erenkov detector to
distinguish electrons from pions on an event-by-event basis. Instead, a
statistical method was used, which is described in detail in \cite{kth}
and is very similar to that described in \cite{yuly}.

\subsection{Acceptance and Dispersion}

The momentum acceptance, $\Delta p/p$, and dispersion were determined by
scanning a $\pi^+ p$ elastic
scattering peak from the CH target across the spectrometer focal plane by
changing the spectrometer magnetic field. The changing acceptance across
the
focal plane resulted in changes in the observed area of the peak. The
relative acceptance as a function of focal plane position was determined by
this method, and was used to correct $N_{\rm det}$. The effective total 
momentum
acceptance was determined to be about 8\% by integration across the entire
focal plane. 

\subsection{Normalization}

The absolute number of incident pions, $N_{{\rm inc}}$ in Eq.~(\ref
{eqn:xsect}), was obtained by comparison of $\pi p$ elastic scattering
measurements from the CH target with known cross sections determined by
interpolation using the energy-dependent phase shift program
SCATPI~\cite{walter}.  In this way,
the effective total solid angle, $\Delta \Omega$, and the effective
target
thickness, $x$, were included in this calibration. This procedure was
repeated at several scattering angles for each setting of the beam
transport
system in order to improve the accuracy of the normalization, as well as to
check for the presence of angle-dependent effects.

\subsection{Background Subtraction}

The pions counted with the empty target were primarily those scattered from
the walls of the target cell. The background was typically only
about 8\%, and never more than 15\%, of the full-target rate, and was
measured to about 30\% accuracy.
The measurements with the empty target were analyzed in the same manner
as those with the full target, and properly normalized background data
were subtracted point-by-point from the full-target data.

\subsection{Corrections}

Corrections (see Eq. (1)) were made to account for other effects that would 
change the
shape and magnitude of the cross section distributions. The number of pions
detected was reduced by decay as they travelled from the target to the
detectors. Approximately 80\% of the pions survived at 200 MeV, while at 10
MeV only 30\% survived. Some of the decay muons traversed the spectrometer,
while others did not. In fact, it is possible for pions that would not have
traversed the spectrometer to decay into muons that could. Since the
detector system could not separate the muons from pions, these effects were
computed by Monte Carlo methods which produced the factor $f_d$ in
Eq.\
(\ref{eqn:xsect}). This procedure is described in greater detail in Ref.\
\cite{wood}. Muon contamination from pion decay in the spectrometer was
simulated with the DECAY TURTLE~\cite{carey} code. The fraction of muon
contamination ranges from 1\% at 10 MeV to 20\% at 200 MeV outgoing pion
kinetic energy. A different Monte Carlo program, MUCLOUD~\cite{wth,kth},
was
used to simulate decays in the scattering chamber. MUCLOUD used the
uncorrected cross sections in an iterative procedure to obtain the
corrections, which were typically 0--30\%. Energy loss and multiple
scattering in the target and in WC0 changed the effective acceptance of the
spectrometer; these were corrected for by the factor $\epsilon_c$, which
was
determined by simulation to be different from unity only for outgoing pion
energies less than 30 MeV. The maximum value of $\epsilon_c$ was about 1.3.

\subsection{ Integrated cross sections}

Integration of the doubly differential cross sections over outgoing pion
energy yielded an angular distribution. This integration was carried out
using the trapezoid rule.  The cross section below the
lowest measured outgoing energy of 10 MeV as found by linear extrapolation
to zero at zero outgoing pion energy. 
Except for several
forward-angle spectra at high incident pion energy, the doubly
differential cross sections were measured up to the kinematic limit for
the DCX reaction, and thus the high-energy end of the integral required no
extrapolation of the data.  The exceptional cases were extrapolated
linearly to zero cross section using the slope determined by the highest
energy pair of cross section measurements. 
The uncertainty in each
differential cross section was calculated as the sum in quadrature of the
uncertainty in each trapezoid's area. The uncertainty in the extrapolation
of the endpoints to zero cross section was estimated to be one-half of the
contribution to the integral from the endpoint regions.
In all cases these uncertainties\cite{kth} were small compared with the 
systematic uncertainties discussed in the next section.

The smooth angular dependence of the differential cross sections allows a
simple method of integration over angle to determine a quantity which we term
the ``total reaction cross section.''  The angular distributions between the 
most
forward and backward angle measurements  were fitted with sums of
Legendre polynomials in $x = \cos \theta_{\rm
lab}$; the distributions were extrapolated linearly in $x$ from the extreme 
angle
measurements to $0^\circ$ and $180^\circ$, using the derivative of the Legendre
polynomial sum at the extremes to define the slopes.  
The uncertainty in the total reaction cross section was determined by fitting 
the Legendre polynomials to the angular distribution plus and minus the
uncertainty at each angle. This uncertainty and that in the
extrapolation~\cite{kth} were less than or comparable to the 
systematic uncertainties discussed in the next section.

\subsection{Systematic uncertainties}

There are systematic uncertainties of three types: those that depend on the
energy of observation, the angle of observation, and the overall
normalization. The uncertainty that depends on the energy of observation
includes contributions from the electron-pion separation procedure, which
were between 2\% and 5\%, together with
contributions from estimates of the corrections for energy loss and multiple
scattering in the target and in the mid-spectrometer wire chamber, and for
effects of pion decay and muon contamination, which contributed an
additional 5\%. The uncertainty in the correction for energy loss and
multiple scattering was estimated to be equal to one-half of the correction
applied. The uncertainty in the corrections for pion decay and muon
contamination was estimated to be one-half of the difference between the
correction derived from the simulation described earlier, and that given by
accounting only for the loss of pions by decay within the spectrometer.

An angle-dependent uncertainty (see Table I) arises from possible
misalignment between the axis of the target and that of the spectrometer,
which will produce changes in both the thickness of the liquid $^4$He seen
by the beam and in the effective acceptance of the spectrometer. These
uncertainties were estimated together by comparison between the observed
angular variation of the differential cross section for $\pi p$ scattering
and that predicted from interpolation of the known cross sections.

Normalization uncertainties are of two kinds; those that are due to slowly
varying changes in the response of the ionization chamber used to monitor
the incident beam flux and the density of the $^4$He target, and those
that involve the determination of the absolute scale of the measured cross
sections by normalization to $ \pi p$ scattering. Variation in the
ionization chamber response, due to changes in ambient temperature and
pressure, was monitored by comparison with the primary proton beam monitor
and a downstream scattering monitor~\cite{kth}. Since these variations
were
observed to occur on roughly the same time scale as did changes of
spectrometer angle, they have been included with the angle-dependent
uncertainty. 

The uncertainty in the overall scale of the cross section contains
contributions from uncertainties in the densities of the $^4$He and CH
targets and in the assumed $\pi p$ scattering cross section.  The
uncertainties in the densities were determined~\cite{kth} to be 0.7\%
and
0.6\% for the $^4$He and CH targets, respectively.  An additional overall
uncertainty of 2\% is assumed to account for the accuracy of the phase
shift prediction of the free $\pi p$ cross section\cite{walter}.
Combining
these uncertainties with the uncertainty in the determination of each
particle's position in the spectrometer focal plane (1\%) and in the
determination of the spectrometer acceptance function (1\%), yields an
overall uncertainty of 2.4\%, independent of incident beam energy and
angle. (The uncertainties due to the determination of electronics
deadtime, spectrometer dispersion, and wire chamber efficiency were
insignificant compared to those listed above.)  This uncertainty has been
combined in quadrature with the angle-dependent uncertainties in Table 
I to yield the final result in the last column for each incident beam.

\section{RESULTS}

\subsection{Doubly Differential Cross Sections}

The doubly differential cross sections measured for the
$^4$He$(\pi^+,\pi^-)$ reaction at incident energies 120, 150, 180, 240,
and 270 MeV are presented as pion energy spectra in Figs. 3--7,
respectively.  The doubly differential cross sections measured for the
$^4$He$(\pi ^{-},\,\pi ^{+})$ reaction at incident energies of 180 and 240
MeV are displayed in Figs.\ 8 and 9.  The bars on the plotted points
indicate the statistical uncertainty plus the systematic uncertainty in
the pion-electron separation procedure and the corrections for pion decay,
energy loss in the target cell, and multiple scattering in the
inter-dipole wire chamber. 

\subsubsection{Contribution from Pion-Induced Pion Production}

The detection of a pion with charge opposite to that of the incident beam
is not a unique signature of DCX at incident energies above about 170 MeV, 
the
threshold
for the pion-induced pion production (PIPP) reactions
$^4$He$(\pi^+, \pi^+ \pi^-)pppn$ and $^4$He$(\pi^-, \pi^- \pi^+)pnnn$.
The magnitude and kinematic range of the contribution of PIPP is severely
limited by the phase space available to this reaction, however, and a
significant  effect is only possible at incident energies well above 200
MeV.

To indicate the effect that this contribution might have on the shape of
the DCX energy spectra, Figs. 6 and 7 illustrate the position of the 
upper
limit of outgoing pion energy from PIPP as well as the shape of the
available phase space for the reaction $\pi^+$ + $^4$He $\rightarrow
^3$He + $\pi^+ + \pi^- + p$, for the spectra measured at incident
energy 240 and 270 MeV.  The phase space 
predictions, whose 
normalization will be discussed later, are
only presented as a guide.  (The shapes of the energy spectra from PIPP in
deuterium\cite{piaset} were found to be very similar to phase space, but
the measurement of Grion {\it et al.}\cite{grion} of PIPP in $^{16}$O
found
significant deviation from such predictions.)

\subsubsection{Comparison with Phase Space Predictions}

In earlier measurements, Wood {\it et al.}\cite{wood} found that a 
general
feature of the doubly differential cross sections for inclusive DCX in
$^{16}$O and $^{40}$Ca was the similarity of the shapes of the energy
spectra to simple predictions based on the available phase space in a DCX
reaction leading to four bodies:  a pion, two knocked-out nucleons, and
residual nucleus.  Figure 10 compares the inclusive spectra from the
$^4$He and $^{16}$O DCX reactions, measured at 25$^\circ$ and incident
energy 240 MeV; the different curves represent phase space energy
distributions based on different numbers of particles in the final state. 
The phase space prediction for the $^{16}$O data assumes four bodies in
the final state, as mentioned above; this curve has been normalized so that
the phase space prediction integrated over the solid angle and energy of
the outgoing pion yields the measured total DCX reaction cross section. 
The three predictions compared with the $^4$He spectrum are similarly
normalized to the total cross section predicted by a simple scaling
law\cite{grametal}.  The dashed curve is the result for a 5-body final
state for the DCX reaction $\pi^+ + ^4{\rm He} \rightarrow \pi^- + p + p +
p + p$, which is in fact the ultimate final state.  The solid and
dot-dashed curves correspond to 4- and 3-body final states which might
result if two or three of the nucleons in the final state recoil as as a
``cluster'' with no internal motion, i.e., $\pi^+ + ^4{\rm He} \rightarrow
\pi^- + p + p + (2p)$ or $\pi^+ + ^4{\rm He} \rightarrow \pi^- + p +
(3p)$.  It is clear that, unlike the case of $^{16}$O, none of the 
distributions even approximately
represents the data, nor provides any insight into the most striking
feature of the doubly differential cross sections for DCX in $^4$He, the
large peak-like structure at high outgoing pion energy. 

\subsubsection{Kinematic Analysis of the Energy Spectra}

In order to gain more quantitative information about the shape of the energy
spectra, especially the centroid energy and width of the high energy peak, a
line shape constructed from a Gaussian curve and a five-body phase space
distribution was fitted to those spectra in which a peak was apparent.  The
centroid, width, and area of the Gaussian were all allowed to vary in the
fitting procedure, but only the magnitude of the phase space distribution was
variable.  The missing mass and four-nucleon excitation energy (defined as the
missing mass less the mass of the four free nucleons in the final state) was
extracted from this analysis.  Examination of the results (see Ref.\cite{kth})  
clearly indicates that the peak does not correspond to a resonance in the
four-nucleon system.  There exists a systematic variation in the missing mass
with scattering angle and incident energy which is well outside the uncertainty
in its determination.\footnote{For example, the excitation energy at 
25$^{\circ}$ varies from
$38.1 \pm 0.6$ MeV at incident energy 150 MeV to $86.1 \pm 0.9$ MeV at 270 MeV;
and at incident energy 240 MeV from $74.4 \pm 0.7$ MeV at 25$^{\circ}$ to $93.8
\pm 2.5$ MeV at 130$^{\circ}$.} The width of the peak varies less dramatically,
but nonetheless sufficiently to contradict the four-nucleon resonance
hypothesis.

Another hypothesis is that the peak results from quasi-free scattering of
the pion with some sort of ``cluster'' within the nucleus.  If the mass of
the cluster remains unchanged by the scattering, there is no choice of
mass which can describe the magnitude and angular variation of the peak
energy, even at a single incident energy.  If one allows the mass of the
cluster to change, corresponding for example to the excitation of the
cluster into a resonant state, it is possible to find a combination of
initial mass and excitation energy which approximately describes the data
at a given incident energy.  It should be emphasized that a similar
variation results if the scattering predominantly occurs on a certain
number of nucleons, and energy is lost to each nucleon individually.  The
quasi-free kinematic analysis presented here is similar in this sense to
earler rapidity analyses of inclusive energy
spectra\cite{mckeown,wood,schum}.
Ref. \cite {kth} presents the results of fitting the kinematic variation
with two-body scattering formulae, assuming an initial cluster ``mass''
and a final ``mass'' found by adding an excitation energy to the initial
mass.  The initial masses are poorly constrained, but nevertheless show a
large variation with incident energy, as do the excitation energies which
are more accurately determined.  An explanation of the high outgoing
energy peak in terms of quasi-free scattering and excitation therefore
seems unlikely.

In Section VII.B a simple double scattering reaction model is
considered; the results suggest that this peak has a natural explanation
in the dynamics of two forward angle $\pi-N$ reactions.

\subsubsection{Comparison of
$^4$He$(\pi^+,\pi^-)4p$ and $^4$He$(\pi^-,\pi^+)4n$ Reactions}

To the extent that Coulomb effects may be neglected, $(\pi^+,\pi^-)$
and $(\pi^-,\pi^+)$ are identical reactions in nuclei with equal numbers
of protons and neutrons, assuming the invariance of the strong interaction
with respect to rotations in isospin space.  Figures 11 and 12 compare 
a sample of
the energy spectra for the two reactions.  The error bars indicated in the
figure do not include the uncertainty in the normalizations which are 6\%
and 7\% for the 180 and 240 MeV spectra, respectively.  As can be seen,
the shapes and magnitudes of the spectra are quite similar.  

This near equality of the cross sections for these mirror processes can be
understood by estimating the sizes of effects caused by the Coulomb
force\cite{kth}.  It is found that the distortion of the pion wave
causes a shift of at most 6 MeV in the $\pi$-nucleus interaction between
the two reactions.  The effect of the Coulomb energy on the distribution
of the nucleons in the $^4$He ground state is estimated to produce a
difference of about 1 MeV between the neutron and proton distributions. 

The energy of the four-nucleon final state is affected more strongly by
the Coulomb potential; the potential energy difference between four
protons and four neutrons is approximately 12 MeV.  Thus, reactions
leading to states of four protons with minimum internal motion could be
suppressed or their energy shifted relative to four-neutron states,
because of the Coulomb energy barrier.  On the other hand, the energy of
the first excited state of $^4$He indicates the effective ``strength'' of
the Pauli principle to be about 20 MeV.  Two nucleons must be excited out
of initial $l = 0$ eigenstates in $^4$He by the DCX reaction, giving a
total energy barrier of approximately 40 MeV for a four-like-nucleon
``ground state.''  Since the Pauli principle applies to both four-proton
and four-neutron final states, the Coulomb potential energy difference
therefore appears to be less significant, though not negligible, in
comparison with the effect of the Pauli principle.  It should be noted 
that
Coulomb effects are expected to be sizable only at the highest outgoing
pion energies, which correspond to little energy left in the residual
system of four nucleons.

\subsubsection{Comparison with DCX in $^3$He}

Figure 13 compares the doubly differential cross sections for the
reactions $^3$He$(\pi^-,\pi^+)$ and $^4$He$(\pi^-,\pi^+)$ measured at
50$^\circ$ and 240 MeV incident energy.  The shapes of the energy spectra
are strikingly similar.  The magnitude of the DCX cross sections in $^3$He
is slightly larger than in $^4$He, and the peak is at a somewhat
higher outgoing pion energy.  The latter difference between the two
spectra may arise simply from the difference in binding energies of the
nuclei, since the DCX reaction leads to complete disintegration of the
nucleus in both cases.

The difference in the magnitude of the differential cross sections may
result from the different strengths of the competing reactions of
quasi-free scattering and pion absorption.  Recalling that negative pions
interact more strongly with neutrons (in the $\Delta$ resonance region),
the competition from quasi-free scattering in  $^4$He should be twice as
large as in $^3$He.  Likewise, if one considers pion absorption
to proceed through a ``quasi-deuteron" mechanism, twice as many $np$ pairs
are found in $^4$He as in $^3$He.  Indeed, recent measurements of pion
absorption cross sections in $^3$He and $^4$He \cite{abs} show the 
latter cross
section to be approximately twice the former, in the $\Delta$
resonance region. Thus, the absorption process is expected to
compete more strongly with DCX in $^4$He.  
However, the DCX reactions should also be sensitive to the
density of the nucleus.  One expects double scattering to be enhanced if
the average internucleon distance is smaller in one case compared to
another, suggesting a larger DCX reaction cross section in $^4$He than in
$^3$He.  Thus, it is not obvious how large a difference, if any, should 
be
expected between the magnitudes of the DCX cross sections, without a 
more quantitative investigation of the problem.

\subsubsection{Comparison with Previous Measurements}

Of the previous experiments\cite{gilly,fal1,fal2,ungar,stetz1,stetz2} which 
have
investigated the DCX reaction in $^4$He in the energy region of the
present experiment, only one\cite{stetz1,stetz2} affords the possibility of 
comparison with 
the
present results for the doubly differential cross sections.  Two 
experiments\cite{gilly,ungar} were searches for
tetraneutron bound states, performed for a limited range of incident and
outgoing pion energies at a scattering angle of 0$^\circ$.  We note,
however, that the magnitudes of the cross sections measured in both of the
search experiments are consistent with the magnitude of the doubly
differential cross section presented here.

Comparison of the present data with those of Stetz {\it et 
al.}\cite{stetz2} is
problematical, since no spectra were measured at a common incident energy
and scattering angle.  
In Ref. \cite{kth} the Stetz {\it et al.} data are compared with 
results of the present experiment at nearly the same
scattering angle but for incident energies above and below, thus
``bracketing" the earlier data.   The comparison of the Stetz data
at 140 MeV with the present data at 120 and 150 MeV (Figs. 3 and 4) 
shows
reasonable consistency in the upper region of the pion energy spectrum,
but the Stetz spectrum has a very different shape at lower pion
energies.
The comparison of the Stetz data at 200 MeV with the present data at 180
and 240 MeV (Figs. 5 and 6) reveals a disagreement in both magnitude 
and shape.  The reason
for this discrepancy is not known, although some possible causes are
suggested in Ref. \cite{kth}.

In the present experiment neither the uncertainty arising from
the normalization of the data to elastic $\pi p$ scattering nor that arising 
from the corrections that depend upon
outgoing pion energy is
sufficient to explain the discrepancies.  

\subsection{Differential Cross Sections}

The singly differential cross sections (angular distributions) 
resulting from integration of the pion energy
spectra as described in Sect.~V.G are displayed in Fig.~14.     
The angular distributions for the $(\pi^+, \pi^-)$ and $(\pi^-, \pi^+)$
reactions are seen to be essentially identical in shape and magnitude
at 240 MeV.  At 180 MeV the differential cross section for $(\pi^+,
\pi^-)$ is slightly larger and more forward-peaked than that for
$(\pi^-, \pi^+)$.

\subsection{Total Reaction Cross Sections}

The total reaction cross sections obtained as described in Sect.~V.G are given 
in the second column 
of Table II and are displayed in Fig.~15.  The
cross sections for the $(\pi^+, \pi^-)$ and $(\pi^-, \pi^+)$ reactions are
seen to be essentially identical at 240 MeV, whereas the former slightly
exceeds the latter at 180 MeV.

The DCX cross section does not exhibit
a resonance shape near the energy of the $\Delta$ resonance, in contrast
to other pion-nucleus reactions.  This can be understood as a simple
consequence of double scattering, if on average the pion loses substantial
energy in its first interaction.  As the incident pion energy moves
through the $\Delta$ resonance, the energy of the second collision is
still below resonance, whereas at incident energies 10--100 MeV higher
than the $\Delta$ resonance, the energy of the second collision is just
passing through the $\Delta$ region.  The effect is to broaden the
resonance and shift its maximum to higher energy.  The total DCX cross
section presumably falls at higher energy, as both collisions occur at
energies above that of the $\Delta$ resonance.  Previous measurements 
at higher energies\cite{car,jean}
do indeed support this hypothesis.

\subsubsection{Contribution of pion-induced pion production}

PIPP is commonly expected to occur on a single nucleon in a quasi-free
manner; the measurements in deuterium performed by 
Piasetzky {\it et al.}\cite{piaset} support this
hypothesis.  In heavier nuclei, the effect of absorption and other
competing $\pi$-nucleus reactions and of the nuclear medium should be
taken into account.  The procedure often followed to estimate the
 nuclear PIPP cross section is to multiply the free PIPP cross section
by an ``effective'' number of protons or neutrons, which can be derived
from analysis of the angular distribution of inclusive quasi-free pion
scattering\cite{kth}.  Precise data on
the $\pi^- p \rightarrow \pi^- \pi^+ n$
cross section have been reported by Bjork {\it et al.}\cite{bjork} in 
the incident energy range 203 -- 357 MeV. The 
observed\cite{piaset} equality of the 
$\pi^+ d \rightarrow \pi^+ \pi^- pp$ and $\pi^- d \rightarrow \pi^- 
\pi^+ nn$ cross sections, and of the $\pi^- d \rightarrow \pi^- \pi^+ 
nn$ and $\pi^- p \rightarrow \pi^- \pi^+ n$ cross sections, justifies 
using the results of Bjork {\it et al.}\cite{bjork} for both positive 
and negative incident pions.  The PIPP cross section at 180 MeV was 
estimated from an extrapolation of the data of Kernel {\it et 
al.}\cite{kernel}, which is consistent with the model-independent fit, 
using four isospin amplitudes, of Burkhardt and Lowe\cite{bl}.  Taking
the effective numbers of nucleons
for $^4$He from Ref.\cite{kth}, one obtains the PIPP cross sections
given in the third column of Table II.  An
alternative estimate of PIPP in $^4$He may be obtained from data on the
$^3$He$(\pi^+, \pi^-)$ process, which cannot proceed via double charge
exchange.  This cross section was measured at 240 MeV incident energy
at two angles (25$^\circ$ and 130$^\circ$) by Yuly {\it et
al.}\cite{yuly}.  If one assumes that the angular distribution of PIPP
in $^3$He is similar to that in $^2$H, one can use the results in
Ref.\cite{piaset} to estimate a total cross section for the former
process. Multiplying this result by two\footnote{Use of this factor
implicitly assumes that absorption, final-state interactions, and Pauli
blocking effects are the same for $^3$He and $^4$He.} to account for
the extra neutron in $^4$He gives the result shown in the fourth 
column
of Table II, which is roughly consistent with the corresponding result 
in the third column.  Subtracting the cross sections given in the 
third
column from the measured $^4$He inclusive cross sections yields the
``corrected'' DCX cross sections given in the fifth column.

The phase space distributions in Figs. 6 and 7 are normalized so as to 
yield the total PIPP cross sections in the third  column of Table II.

\subsubsection{Comparison with other measurements}

Falomkin {\it et al.}\cite{fal1,fal2} have reported total reaction cross
sections in the energy region between 98 and 156 MeV.  However, these
measurements are
statistically inaccurate, and may also suffer from an unknown systematic
error (see Ref.\cite{kth}).  However, within their limited accuracy, the
magnitudes of these total cross sections are in overall agreement with
those derived from the present data.  The more accurate total cross sections 
reported by Stetz {\it et 
al.}\cite{stetz1}  appear to be systematically lower than the present results by 
approximately 30-50\%.  This difference is greater than the 
uncertainties arising 
from normalization and integration in both experiments (see Ref. \cite{kth}). 

The total cross section resulting from the more recent measurement of 
Gr{\" a}ter {\it et al.}\cite{grater} is in good agreement with 
the 
present results where the two data sets overlap at 120 MeV.

\subsubsection{Comparison with theoretical calculations}

Several theoretical predictions for the total cross section have been
reported.  The prediction of the sequential single charge exchange calculation 
of Becker and
Schmit\cite{bs} (the region between the two curves labeled B-S in Fig.~15) is 
clearly of the wrong
magnitude, and rises more rapidly with incident energy than does the
measured total cross section.  This is presumably because of the lack of
any treatment of the effect of pion absorption, through distorted waves
for example, as well as the neglect of binding effects and the use of the
fixed nucleon approximation.  The prediction of Germond and
Wilkin\cite{gw} (curve labeled G-W in the figure), based on pion
scattering from virtual pions exchanged by the nucleons (see Fig.~1(b)), has an 
energy
dependence more similar to that of the data, but is low by a factor of
approximately three.  Recalling that this calculation does not include the
effects of important cancelling terms in the reaction amplitude, the real
contribution from such meson exchange mechanisms for DCX is expected to 
be
even smaller.  The prediction of H{\" u}fner and Thies\cite{ht}(curve labeled 
H-T in Fig.~15) also
assumed a multiple scattering type of DCX reaction mechanism.  The cross
sections are calculated with a cascade approach to pion-nucleus scattering
wherein the pion propagates classically between collisions.  The effects
of the competition of the DCX reaction with pion absorption and inelastic
scattering are included in the calculation, and these effects are expected
to be critical to obtaining quantitative understanding.  Indeed, the
results of the calculation are in relatively good agreement with both the
magnitude and the energy dependence of the cross sections presented here. 
Below 200 MeV incident pion energy, the calculation lies 10--30\% higher
than the measured cross sections, whereas it begins to fall below them at
higher incident energies.  Some of this discrepancy at higher energy may
be the result of the contributions to the measured cross section from
pion-induced pion production, as suggested in the figure, but the 
theoretical prediction appears
nonetheless to be low. 

\section{Comparison of Doubly Differential Cross 
Sections for DCX with Theoretical Models}

\subsection{Fixed-Nucleon Sequential Single Charge Exchange Model}

With the aim of calculating doubly differential cross sections with which
the present data could be compared, Rebka and Kulkarni\cite{rebka,kul}
reviewed and modified the fixed-nucleon SSCX calculation of Gibbs {\it et
al.}\cite{gibbs}, as described in Section III.  The results for
$^4$He($\pi^+,\pi^-$) at incident energy 240 MeV and outgoing pion angles
$25^{\circ}$ and $50^{\circ}$ are shown in comparison with
the data in Fig. 16.  More recently, Alqadi and
Gibbs\cite{alqadi} have performed a fixed-nucleon calculation and
investigated the effect of inclusion of final state interactions (FSI).
Although differing somewhat in magnitude, the two
calculations\cite{kul,alqadi} without FSI yield qualitatively 
similar results (compare Fig. 16 with Fig. 1 in Ref.\cite{alqadi}).  
The effects of FSI are seen \cite{alqadi} to be largest for high pion 
energies.  This is
expected, as pointed out by Alqadi and Gibbs, since in this region the
relative momentum of the two final state nucleons is smallest.  None of 
these calculations reproduce the magnitude or the shape of the doubly
differential cross section.  The discrepancies between the data and the
modified Gibbs calculation are found to persist at other angles 
and incident energies \cite{kul}.

\subsection{Non-Static Sequential Single Charge Exchange Model}

At the time of the completion of the present experiment, there were no 
published theoretical doubly differential cross
sections or angular distributions for DCX in $^4$He with which the 
data could be compared.
However, several authors had performed
simple calculations in order to gain some understanding of the general
features of the data.  Wood\cite{wth} developed a simple cascade model to
describe DCX in $^{16}$O.  The scattering probabilities per unit volume
were derived by averaging the free $\pi N$ cross sections over a Fermi gas
nuclear momentum distribution.  While the doubly differential cross
sections were similar in shape and magnitude to the data, the angular
distribution and the energy dependence of the total cross section were not
well described\cite{wood}.  A simpler calculation, also described in Ref.
\cite{wth}, consisted of a folding of two free SCX cross sections,
assuming a Fermi gas model of the nucleus.  Pauli blocking and Fermi
motion effects were investigated and found to be important.  A similar
folding calculation by Thies\cite{tpc}, using Gaussian wavefunctions to
describe $^4$He, found very similar results.  Van Loon\cite{vl} 
attempted to include the medium effects by parameterizing the energy
spectra from inclusive quasi-free scattering and using the results in a
folding calculation to obtain DCX spectra; the inclusive scattering was
corrected by the ratio of the free $\pi N$ cross sections to obtain the
effective inclusive SCX spectrum, since no inclusive SCX data were 
available for $^4$He.

The major finding of Wood, Thies, and van Loon was that a high energy peak
and a low energy peak in the forward angle DCX energy spectrum result
from the $(1 + 3 \cos^2 \theta)$ angular dependence of the $\pi N$ cross
section near the $\Delta$ resonance.  This shape leads to enhanced forward
and backward scattering.  Two forward scatterings yield a forward-going
pion of high energy, and two backward scatterings also yield a
forward-going pion, with much lower energy.  At larger angles, the favored
combinations of individual forward and backward scatterings are not as
clearly separated kinematically, hence the double-peaking feature is not
as clear.  The cascade calculation of Wood suggests that
absorption and higher-order multiple scattering obscure these features in
$^{16}$O; the comparison of the simpler folding models with the $^{3,4}$He
data is however more favorable, and suggests that development of the
folding
calculations would be worthwhile.  With this in mind, and also in the
belief that proper treatment of the Fermi motion and binding effects
would be important, we undertook a calculation based on the SSCX mechanism.  

This treatment is an extension of some of the work of van Loon\cite{vl}
and relies on the formalism developed by Thies\cite{thies} to describe
inelastic, multistep reactions with quantum-mechanical transport theory.  
A number of approximations and assumptions are made, some of which are
known to be quantitatively inaccurate in the precise description of
pion-nucleus scattering, especially in the energy region of the $\Delta$
resonance;  this calculation in no way aspires to be an exact description
of pion double scattering, but its qualitative features may elucidate the
important physics of the reaction.

A brief summary of the essential features of the calculation is given in
Ref. \cite{yuly}; an extended description is contained in Ref.
\cite{kth}.  The technical details
will be published elsewhere\cite{kpap}.

In this model the incident pion interacts sequentially with two
like-charge nucleons only, and thus
only the leading, or ``double scattering'', term in the transition
matrix is used, {\it i.e.},
\begin{equation}
T=\sum_{i=1}^A\sum_{j\neq i}^At_iG_0t_j,
\end{equation}
where the $t_i$ are the in-medium transition operators ($t$-matrices) for
scattering from
the $i$th nucleon, and $G_0$ is the in-medium pion propagator. 
What distinguishes this calculation from simpler folding models is the
non-static treatment of
the $\pi N$ interaction and the inclusion (albeit approximate) of
the binding of the nucleons. 

In the energy region of the $\Delta$ resonance, the interaction between
the pion and the nucleon varies strongly with pion-nucleon relative
energy. Because of the the Fermi motion of the bound nucleon, the
treatment of the interaction within the nuclear medium is a critical part
of any calculation of pion-nucleon scattering.  In particular, one wishes
to account for the variation in interaction energy due to simple
kinematics and the mean nuclear potential, both of which are well known.  
Following the treatment by Lenz\cite{lenz}, we replace the in-medium
$t$-matrix by the free $\pi N$ $t$-matrix evaluated at an effective
relative energy:  \begin{equation} t_i(E)=t_{{\rm free}}(E-T_{{\rm
c.m.}}-U({\bf r}_i)-H_{A-1}), \label{eqn:inmedt} \end{equation} where
$H_{A-1}$ is the nuclear Hamiltonian with the dependence on the $i$th
nucleon separated out, $T_{{\rm c.m.}}$ is the center-of-mass kinetic
energy of the system formed by the $i$th nucleon and the incident pion,
and $U({\bf r}_i)$ is the mean nuclear (shell-model)  potential felt by
the $i$th nucleon.  As discussed in Ref. \cite{kth}, the dependence of the
shell-model potential on ${\bf r}_i$ makes a general solution
intractable; here the shell model potential will be replaced with a 
constant, $U_i$.  Three choices of $U_i$ were investigated, -55 MeV, 0 
MeV, and -37 MeV, corresponding to values of $U({\bf }r_i)$ in the nuclear 
interior, the surface, and the wavefunction-averaged value 
$(\langle ^4{\rm He} | U({\bf r}_i) | ^4{\rm He} \rangle)$, 
in which harmonic oscillator functions fitted to the 
elastic charge form factor were used.  To avoid this approximation, a full 
$\Delta$-hole calculation \cite{Dh,HLY} would be necessary, which is 
beyond the scope of 
the present work.

Proper treatment of the off-shell
momentum dependence requires a model of the $\pi N$ interaction;
typically, the off-shell effects are described with form factors which are
functions of the relative momentum, and the individual $\pi N$ partial
waves are treated separately\cite{HLY,lenz}.  In this work, these
off-shell
form
factors are not included; the free $t$-matrix is calculated given the
effective relative energy and the scattering angle in the center-of-mass
system.  

This modified free $t$-matrix is inserted into Eq. (2) in order to 
calculate the exclusive DCX amplitude.  The undetected nucleon momenta 
are then integrated over to obtain the inclusive cross section.
To simplify the computation, however, the intermediate pion is
constrained to propagate classically through the nucleus.  The expression 
for this propagator is found using a
Wigner transformation~\cite{thies,kth,kpap} and taking the classical limit
$\hbar \rightarrow 0$.   
Selecting the specific amplitude for scattering from nucleon 1 to nucleon
2, as is shown in Fig. 17, we may write
\begin{eqnarray}
A_{f0}(\vec{k}^{\prime },\vec{p}_1^{\;\prime },\vec{p}_2^{\;\prime
},\vec{k}
) &=&\int d^3q\int d^3p_2\;\langle \vec{k}^{\prime },\vec{p}_2^{\;\prime
}|\,t_2\,|\vec{q},\vec{p}_2\rangle \langle \vec{p}_2|\phi _{1s}\rangle \\
&&\times \frac 1{E-q^2/2m_\pi -(p_1^{\;\prime })^2/2M_N-E_1}\int
d^3p_1\langle \vec{q},\vec{p}_1^{\;\prime }|\,t_1\,|\vec{k},\vec{p}
_1\;\rangle \langle \vec{p}_1|\phi _{1s}\rangle ,  \nonumber
\end{eqnarray}
where $E_1$ is the energy of nucleon 1. This amplitude is calculated
numerically. 

Several further simplifications have been made.  Within a distorted wave
impulse approximation (DWIA) approach, Thies\cite{th2,baum} has shown that
the
inclusive quasi-free $(\pi, \pi ')$ reaction may be largely
explained
with a one-nucleon knockout mechanism, where the residual nucleus is
described by the appropriate hole state.  Making the same assumption here,
the sum over intermediate nuclear states is restricted to a single state.
The further approximation is made of describing the initial state nucleus 
with the shell
model; since the final state of $^4$He after DCX is four free nucleons,
the shell model may not be an apt description.  However, to do better one
would need to solve the four-body nuclear Hamiltonian exactly.
Finally, an enormous calculational simplification is gained by replacing
all of the
distorted wave continuum states (nucleon and pion) with plane waves.

The general result can be further simplified by considering its
application to the specific case of $^4$He.  Since the final nuclear 
states
will involve large excitation energies, the effect of the overall
antisymmetrization of the four-nucleon states is presumed small, except for
the lowest nucleon energies. By requiring that the momentum of the struck
nucleon lie outside the Fermi surface of the ground state, one effectively
includes Pauli blocking of the low energy nucleons, which is the major
effect of the antisymmetrization.  The ground state of $^4$He is assumed
then to be described by the product of four ($1s$) harmonic oscillator
wavefunctions. 

In the calculation of the inclusive cross section, corrections have been
included for the effects of nuclear recoil and center-of-mass motion. 
Applying the latter correction to the harmonic oscillator wavefunctions
for $^4$He, it is found\cite{kth} that the size parameter $1/b^2$ must be
replaced by $4/(3b^2)$. The value $b$ = 1.36 fm was determined from low
momentum transfer electron scattering data \cite{frosch}.
The free transition matrix element used in
the present calculation was derived from the free $\pi N$ cross section
predicted by Arndt~\cite{arndt}, which is an empirical energy-dependent
phase-shift representation of the cross sections.

Figure 18 compares the results of the calculation with the
DCX energy
spectrum measured at 25$^\circ$ for 240 MeV positive pions.  The solid
curve corresponds to an average nuclear potential equal to $-55$ MeV for
both scatterings, i.e. $U_1 = U_2 = -55$ MeV.  The dot-dashed curve
corresponds to scattering with zero potential, $U_1 = U_2 = 0$, and the
long
dashed curve corresponds to $U_1 = U_2 = -37$ MeV, the expectation value
of the potential.  All three of these curves resemble the shape of the
energy spectrum, while the reproduction of the position of the high energy
peak possibly favors the full potential of $-55$ MeV.  As a further
investigation of the effect of the different potentials, the dotted curve
corresponds to $U_1 = -55$ MeV, $U_2 = 0$, and the short-dashed curve
corresponds to $U_1 = 0$, $U_2 = -55$ MeV.  The shape and magnitude of the
prediction are evidently sensitive to the choice of potential for each
scattering; the critical nature of the $\pi N$ $t$-matrix energy
prescription is therefore underscored.  This sensitivity is further
illustrated by the double-dot-dashed curve, which assumes the nucleons to
be
at rest when calculating the interaction energy (static prescription), and
$U_1 = U_2 = 0$. The static prescription fails to reproduce
the low-energy shape of the spectrum, but the position of the high-energy
peak agrees relatively well with the data.

One can understand the difference between the static and non-static
calculations by considering the difference between the shape of the free
$\pi N$ angular distribution in the laboratory and center-of-mass
(c.m.) frames. 
The symmetric $(1 + 3 \cos^2 \theta)$ c.m. angular distribution at
$\Delta$ resonance energies becomes forward-peaked in the laboratory.  The
cross section at forward laboratory angles resulting from the static
prescription ``uses" the laboratory angular distribution
in every collision, and thus only two forward scatterings are preferred
and only the high energy peak results.  

The non-static prescription includes the effect of the Fermi motion of
the nucleons, and also, as mentioned previously and discussed more fully
by Thies\cite{th2,fth}, the strong energy dependence of the $\pi N$
interaction.  This strong energy dependence leads to a preference
for the $\pi N$ interaction to occur with nucleons such that the relative
energy is near the energy of the $\Delta$ resonance.   In double 
scattering, the
enhancement of one frame over another will affect the relative size of
the low and high energy parts of the DCX spectra at forward angles. 

Figure 19 shows a comparison of the calculation with the data for 240 
MeV
incident energy at scattering angles 25$^\circ$, 50$^\circ$, 80$^\circ$,
105$^\circ$, and 130$^\circ$, for the three values of $U_1 = U_2$.  It is
seen that, as the scattering angle increases, the shape as well as the
magnitude of the predictions becomes less similar to the measured cross
sections.  The chief defect in the shape appears to be that the calculated
spectra are concentrated in too narrow a range of outgoing pion energy.
The choice of the value of the potential does move the predictions over
the broader range of the data, but there appears to be no one choice, even
allowing $U_1 \neq U_2$ (see Ref. \cite{kth}) that produces agreement.

The variation in the shape of the outgoing pion energy spectrum with
incident pion energy is fairly well reproduced by the calculation, as
illustrated in Fig. 20, which shows the 25$^\circ$ doubly differential
cross section at 120, 150, 180, 240, and 270 MeV. Calculated cross
sections are shown for three choices of the potentials $U_1 = U_2$.  
In each case the theoretical predictions has been renormalized to yield
the same integrated area as that of the measured energy spectra.  The
inclusion of a non-zero potential appears to be important to
determining the correct shape in most of the spectra. The comparisons
shown in Fig. 20 disguise the fact that the calculation is not able to
reproduce the incident energy dependence of the magnitude of the cross
section -- see Fig. 21.  Only the prediction with $U_1 = U_2 = 0$ has
an energy dependence at all similar to that of the data.  The inclusion
of distortion of the pion waves is expected to cause a large reduction
in the cross section, especially below the energy of the $\Delta$
resonance where pion absorption is an important channel, along with
quasi-free scattering.

\subsection{DCX Cross Section using Relativistic Three-Body Model}

The qualitative and semi-quantitative success of the calculation described 
above in reproducing the data indicates that nucleon motion and 
nuclear binding effects are important, although these effects were 
included in a somewhat {\it ad hoc} fashion.  Ideally, the 
calculation should include a proper treatment of the initial nucleon 
motion and nucleon recoil, along with use of relativistic pion and nucleon 
kinematics, a full integration over the intermediate pion momentum, and 
incorporation of nuclear binding by means of a model which would also fix 
the energy arguments of the $\pi N$ $t$-matrix elements.  Progress 
toward these goals was made by
Kulkarni\cite{kul}, who developed a relativistic three-body model of 
pion-nuclear interactions -- the three bodies being the pion, the $i^{\rm 
th}$ nucleon, and the residual nucleus or ``core'' -- in order to take 
account of 
medium modifications of $\pi N$ interactions.  He has calculated doubly 
differential cross sections for DCX based on the SSCX mechanism, with the 
$\tau$-matrix elements for the first scattering obtained from the 
three-body model.  An attractive square well potential is used to model 
the interaction between the $\pi N$ composite and the core.  The energy 
arguments  of the $t$-matrix elements entering the approximate expression 
for the $\tau$-matrix elements are fixed by means of this prescription, 
which reduces to the prescription of Garcilazo and Gibbs\cite{gg} in 
the non-relativistic limit.

Various investigations were performed\cite{kul}:  cross sections 
were calculated with and without the medium-dependent terms, with 
different widths of the the initial gaussian nuclear wave function, and 
with only $s$-waves and only $p$-waves in the $\pi N$ interaction.  The 
relative contributions of non-spin flip, single- and double-spin-flip 
processes to the cross section were calculated.  Somewhat 
surprisingly, the inclusion of medium-dependent terms was seen to have 
only a small effect on the cross section.  The double-peaking in the 
forward-angle cross sections was confirmed to arise from the $p$-wave 
component of the $\pi N$ interaction, as expected from previous 
work\cite{wth}.

Reference\cite{kul} contains comparisons of the relativistic three-body 
model calculation with a large number of the cross sections measured in 
the present 
experiment.  In general, the agreement is poor, with the exception of the 
case at 120 
MeV, where the theory reproduces both the shape and magnitude of the 
measurements.  A typical result of this calculation is shown in Fig. 
22, which can be compared with Fig. 19.  Although this 
calculation produces double peaks at forward angles, they do not 
resemble those seen in the data, and the disagreeement with the 
large-angle cross sections becomes even greater than that seen in Fig. 
19.

\subsection{Intranuclear Cascade Calculation with Inclusion of Pion 
Production}

Alqadi and Gibbs\cite{alqadi} have calculated doubly differential cross 
sections for $^4$He($\pi^+,\pi^-$) using an intranuclear cascade (INC) 
code.  Pions are allowed to propagate classically through the nucleus and 
interact with the nucleons via absorption, elastic scattering, charge 
exchange, or pion production processes as permitted by energy and charge 
conservation, with the relative interaction probabilities given by 
the ratio of the relevant cross sections to the total cross section.  
The effects of Pauli blocking and final state interactions are taken into 
account.
Comparisons of the calculations with the present data at 240 and 270 
MeV are shown
in Ref. \cite{alqadi}; the results will be summarized here.

1.  When only two nucleons (neutrons) are allowed to be active in the INC, 
the resulting $(\pi^+,\pi^-)$ cross section is found to be very similar to 
that obtained 
using the two-nucleon SSCX model discussed in Sect. 
VII.A.\footnote{See Fig. 2 in Ref. \cite{alqadi}}

2.  Allowing all four nucleons to be active in the INC reduces the cross 
section by a factor of $\sim$2, in better agreement with the data.
The two protons, with their 
large cross section for elastic scattering of positive pions, effectively 
``shield'' the neutrons from the incoming pions and thus reduce the 
probability 
of their initiating of a DCX reaction.\footnote{See Fig. 3 in Ref. 
\cite{alqadi}}

3.  Final state interactions, as in the two-nucleon model, mainly affect 
the cross section for outgoing pions at forward angles and high energies.

4.  At 180 and 240 MeV, as one goes from forward to backward angles, the 
calculation goes from being slightly lower than the measured cross 
section (at 25$^{\circ}$ and 50$^{\circ}$) to being slightly larger (at 
105$^{\circ}$)  to being as much as a factor of two larger (at 
130$^{\circ}$).\footnote{See Figs. 4 and 6 in Ref. \cite{alqadi}}  
This angular dependence, though not as severe here, is 
reminiscent of that found in the non-static SSCX calculations described 
above.

5.  At 240 and 270 MeV, the calculation with pion production ``turned 
off'' does not exhibit a double-peaked structure; the low-energy peak is 
absent.  Inclusion of pion production adds strength to the low pion energy 
region, as expected, so that the resulting theoretical cross section is 
double-peaked, in fair agreement with the data at 25$^{\circ}$ and 
50$^{\circ}$.\footnote{See Figs. 7 and 8 in Ref. \cite{alqadi}}

6.  The calculation is unable to reproduce the double-peaked structure 
seen at 25$^{\circ}$ and 50$^{\circ}$ for incident energy 180 
MeV.\footnote{See Fig. 6 in Ref. \cite{alqadi}}  The 
low-energy peak in this case cannot be attributed to pion production.

\section{Summary and Conclusions}

The results of an experimental investigation of the pion double charge
exchange (DCX) reaction in $^4$He at intermediate incident energies have
been presented.  Study of the reaction is motivated by interest in
understanding the reaction of pions with the nucleus, particularly the
role of  multiple scattering in such reactions.  Pion double charge
exchange is an ideal probe of such processes because two nucleons must be
involved in the reaction in order to conserve charge.  Moreover, the
intimate connection of the pion to the nuclear potential at long range
makes an understanding of pion-nucleus reactions important to all attempts
to investigate the structure and dynamics of the nucleus.

The measurements presented here constitute the first systematic
investigation of the DCX reaction in $^4$He for a broad range of incident
pion energies and outgoing pion angles.  The doubly differential cross
sections for the inclusive reaction $^4$He$(\pi^+, \pi^- )4p$ have been
measured at five incident pion energies in the range 120--270 MeV, and for
the reaction $^4$He$(\pi^-, \pi^+ )4n$ at 180 and 240 MeV.  At each
incident energy, the cross sections were measured at from three to five
angles between 25$^\circ$ and 130$^\circ$, over the entire energy range
of the outgoing pions above 10 MeV. The ability to detect low energy pions
was particularly important in view of the large fraction of the yield
which results from reactions leading to pions with energies below 50 MeV.
The doubly differential cross sections have been integrated to obtain both
angular distributions and total reaction cross sections. 

An interesting feature of the DCX energy spectra measured at forward
angles is a prominent peak at high outgoing pion energy, 
which
changes in magnitude and energy as the scattering angle is increased; this
feature was apparent in the spectra from both $^4$He$(\pi^+,\pi^-)4p$ and
$^4$He$(\pi^-,\pi^+)4n$ reactions.  Empirical analysis of the peak
indicates that it does not arise from a resonant state of four nucleons or
from a reaction with a ``cluster" in the nucleus.  The structure had not
been observed in earlier measurements, primarily owing to the incomplete
survey of the reactions carried out.  

The results of several calculations of the DCX cross sections based on the 
sequential single charge exchange model of the reaction are presented.
Although calculations employing the fixed-nucleon 
approximation\cite{rebka,kul,alqadi} do not 
reproduce the data, a non-static SSCX calculation is able to account for 
many features of the measurements.
In
this model\cite{kth}, the high energy peak is a manifestation of the basic 
$p$-wave
character of the $\pi N$ interaction near the energy of the $\Delta$
resonance.  Because forward and backward scattering are preferred relative
to scattering at intermediate angles, forward double scattering consists
mostly of two forward or two backward reactions.  The energy losses of
the pions in the two cases are quite different, however, and lead to high-
and low-energy components of the outgoing pion spectrum at forward angles.
In this calculation, all continuum
states are treated as plane waves and precise agreement of the absolute
magnitude of the theoretical predictions with the measured cross sections
is therefore not expected.  The shapes of the predicted energy spectra may
be more meaningful. In particular, to reproduce the shape of the measured
spectra, it was found to be important to treat the $\pi N$ interaction
non-statically so as to include the Fermi motion of the nucleons in the
determination of the relative energy of the $\pi N$ system.  Sensitivity
to the inclusion of the nuclear binding potential was also found.

An attempt\cite{kul} to remove some of the approximations in the 
above-mentioned calculation, by use of a relativistic three-body model of 
$\pi$-nuclear interactions, did not result in improved agreement with the 
data.

A recent intranuclear cascade calculation\cite{alqadi}, with the inclusion 
of the pion-induced pion production (PIPP) process, provides good 
agreement with some of the data at 240 and 270 MeV.  Without PIPP, the 
peak at low 
pion energies in the forward-angle data is missing.  However, PIPP cannot 
be responsible for the low-energy forward-angle peaks seen at 180 MeV or 
at 150 MeV 
(well below threshold).  We note that 
in the $^3$He($\pi^-,\pi^+$) reaction\cite{yuly}, forward angle double 
peaks were seen at all incident energies at which measurements were made: 
240, 180, and 120 MeV.  

In summary, the calculations presented here are seen to reproduce some
features of the measured cross sections.  However, a more complete
theoretical treatment is evidently required to account for the 
shapes, magnitudes, and incident pion energy dependence of the 
doubly differential, singly differential, and total reaction 
cross sections for inclusive DCX in $^4$He.

\section{ACKNOWLEDGMENTS}

This work was supported in part by funds provided by
the U.S. Department of Energy.
We would like to thank S. A. Wood for sharing his wisdom and experience in
the analysis of the data, and W. R. Gibbs for making his code available
for our perusal.  One of us (ERK) wishes to acknowledge E. J. Moniz
and M. Thies for their helpful discussions of the theory.  We also wish 
to thank T. Akdo\u{g}an for his assistance in preparing the figures.

\clearpage



\begin{table}[tbp]
\caption{Systematic uncertainties in the experiment.}
\label{table:TableI}
\begin{ruledtabular}
\begin{tabular}{c|ccc|c}
\multicolumn{5}{c}{Systematic uncertainties (\%)} \\ \hline
& \multicolumn{3}{c|}{Angle Dependent} & \\ \cline {2-4}
Incident Beam & Thick.$^a$ & I.C.$^b$ & Total & Overall$^c$ \\ \hline
120 MeV $\pi^+$ & 5.2 & 1.0 & 5.3 & 5.7 \\
150 MeV $\pi^+$ & 5.5 & 1.0 & 5.6 & 6.1 \\
180 MeV $\pi^+$ & 2.4 & 1.0 & 2.6 & 3.6 \\
180 MeV $\pi^-$ & 4.7 & 2.0 & 5.1 & 5.7 \\
240 MeV $\pi^+$ & 2.3 & 1.0 & 2.5 & 3.5 \\
240 MeV $\pi^-$ & 7.4 & 1.5 & 7.6 & 8.0 \\
270 MeV $\pi^+$ & 7.7 & 1.5 & 7.8 & 8.2 \\
\end{tabular}
\end{ruledtabular}
\footnotetext{$^a$Uncertainty in target thickness due to possible
misalignment (see text).\\$^b$Normalization uncertainty due to variation in
ionization chamber response (see text).\\$^c$Includes angle
dependent and normalization uncertainties only. The uncertainties which
depend on the energy of observation are included along with the statistical
uncertainties in the plotted error bars.}
\end{table}



\begin{table}[tbp]
\caption{Total reaction cross sections for $^4$He$(\pi^+,
\pi^-)$ and $^4$He$(\pi^-, \pi^+)$, and estimated cross 
sections for pion-induced pion production (see text).}
\label{table:TableII}
\begin{ruledtabular}
\begin{tabular}{c|c|cc|c}
\multicolumn{5}{c}{Reaction cross sections ($\mu$b)} \\ \hline 
Incident Beam & $\sigma_{\rm tot}$ & $\sigma_{\rm PIPP}$ & 
$\sigma_{\rm 
PIPP}(^3$He) & $\sigma_{\rm DCX}$ \\ \hline
120 MeV $\pi^+$ & $\; \; 128 \pm \; \; 13 $ & & & \\
150 MeV $\pi^+$ & $\; \; 280 \pm \; \; 18 $ & & & \\
180 MeV $\pi^+$ & $\; \; 489 \pm \; \; 23 $ & $\; \; \; \; 2 \pm \; \; 
1 $ & & \\
180 MeV $\pi^-$ & $\; \; 418 \pm \; \; 25 $ & $\; \; \; \; 2 \pm \; \; 
1 $ & & \\
240 MeV $\pi^+$ & $\; 1014 \pm \; \; 36 $ & $\; \; 128 \pm \; \; 8 $ & 
$\; \; 90 \pm \; \; 20 $ & $ \; 886 \pm 138 $ \\
240 MeV $\pi^-$ & $\; 1075 \pm \; \; 76 $ & $\; \; 149 \pm \; 15 $ & & 
$\; 926 \pm \; 77 $ \\
270 MeV $\pi^+$ & $\; 1460 \pm \; 105 $ & $\; \; 509 \pm \; 50 $ & & 
$\; 951 \pm 116 $ \\
\end{tabular}
\end{ruledtabular}
\end{table}

\clearpage

\begin{figure}[p] 
\begin{center}
\epsfig{file=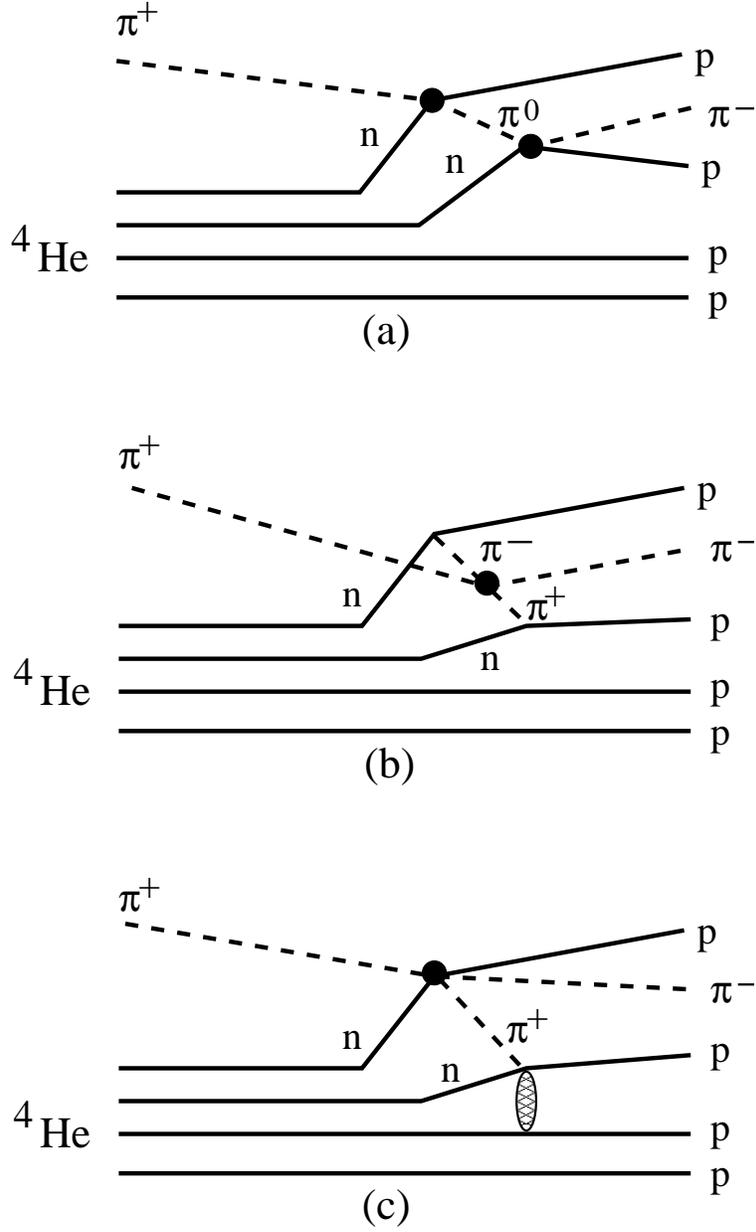,width=0.6\linewidth}
\end{center}
\caption{Schematic diagram of (a) the SSCX mechanism, (b) the
meson-exchange
mechanism of Germond and Wilkin~\protect\cite{gw}, and (c) the
$pn$ absorption mechanism of Jeanneret {\em et al.}~\protect\cite{jean}
for the $^4$He$(\pi^+, \, \pi^-)4p$ reaction.}
\label{fig1}
\end{figure}

\begin{figure}[p] 
\begin{center}
\epsfig{file=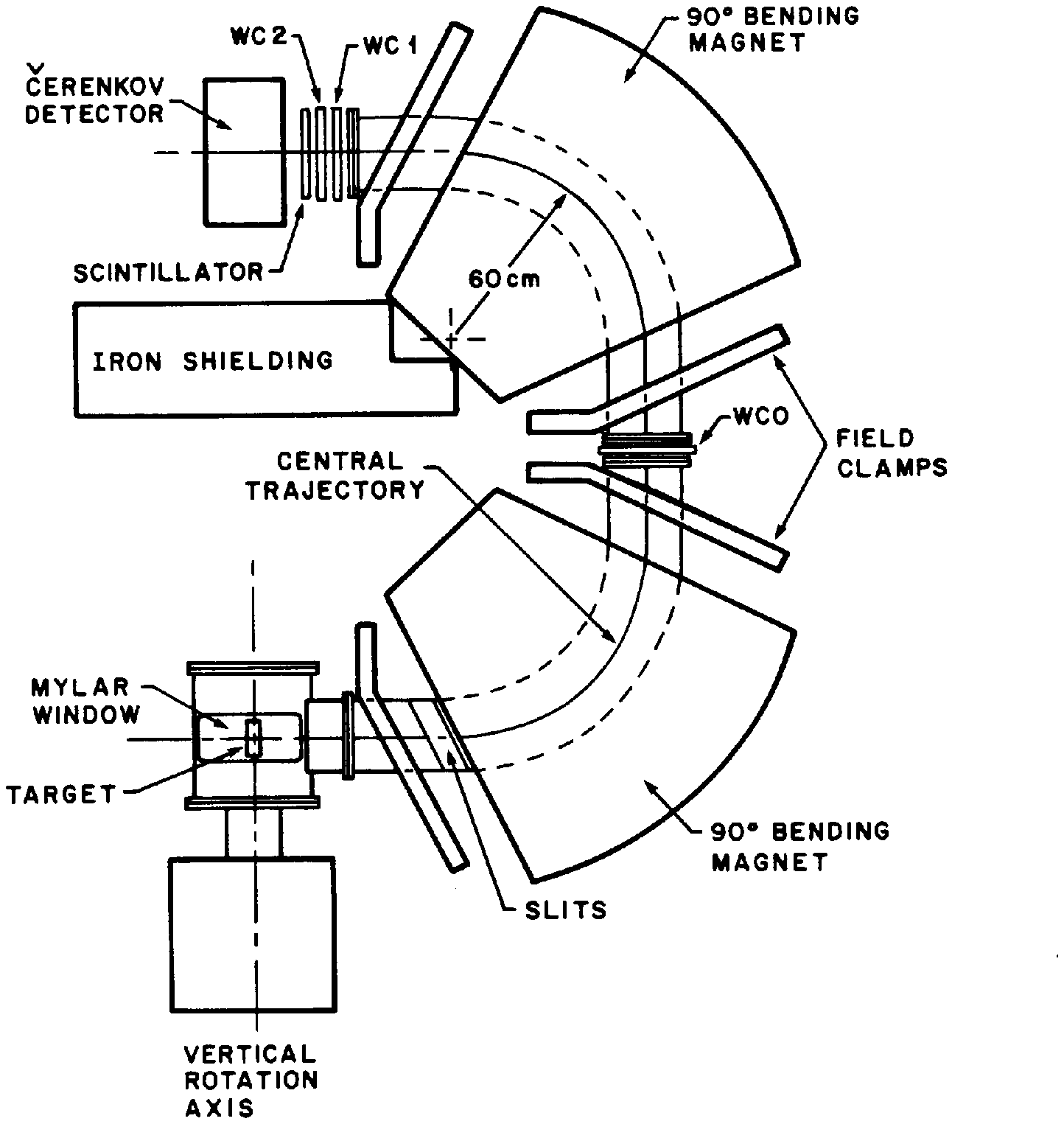,width=0.7\linewidth}
\end{center}
\caption{Drawing of the 180$^\circ$ vertical bend, double focussing
magnetic spectrometer. Pions travel through vacuum from the target, through
two 90$ ^\circ$ dipole magnets, to the focal plane. There is a 2.5
cm break in the vacuum for WC0, the mid-spectrometer wire chamber, which is
used to ensure that a particle traverses the entire spectrometer.
Particle trajectories are traced back to the focal plane using information
from two wire chambers, WC1 and WC2. The scintillator is used to
distinguish positive 
pions from protons, as well as to provide time-of-flight information. The \v 
Cerenkov
detector separates pions from electrons and positrons.  The liquid $^4$He 
cryostat is not shown.}
\label{fig2}
\end{figure}

\begin{figure}[p] 
\begin{center}
\epsfig{file=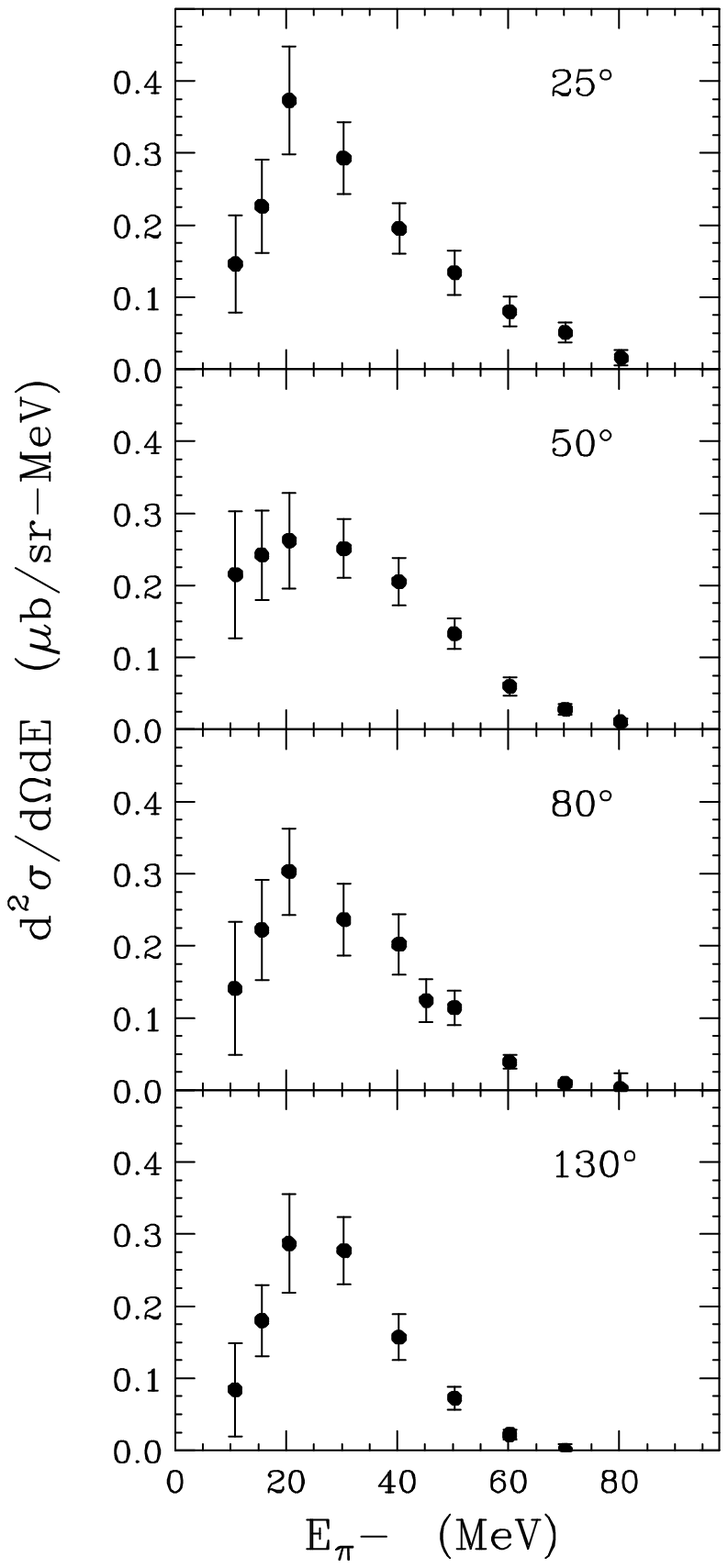,width=0.5\linewidth} 
\end{center}
\caption{Doubly differential cross sections for $^4$He$(\pi^+ ,\, \pi^- )$
at 120 MeV for laboratory angles 25$\protect^\circ$, 50$^\circ$,
80$^\circ$, and 130$^\circ$. 
The uncertainties shown
include the statistical uncertainty and the systematic uncertainties which
depend on the outgoing pion energy.}
\label{fig3}
\end{figure}

\begin{figure}[p] 
\begin{center}
\epsfig{file=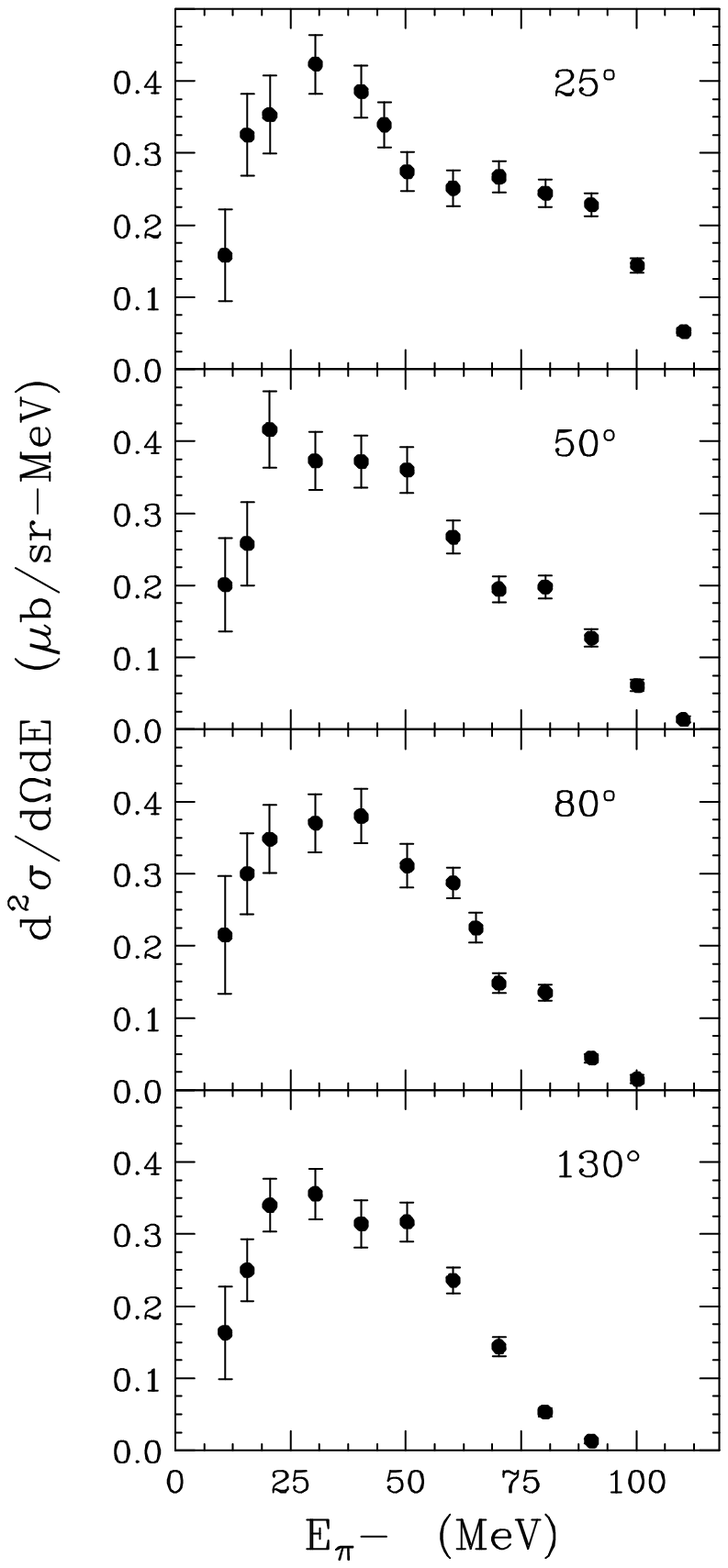,width=0.5\linewidth}
\end{center}
\caption{Doubly differential cross sections for $^4$He$(\pi^+ ,\, \pi^- )$
at 150 MeV for laboratory angles 25$\protect^\circ$, 50$^\circ$,
80$^\circ$, and 130$^\circ$. 
The uncertainties shown
include the statistical uncertainty and the systematic uncertainties which
depend on the outgoing pion energy.}
\label{fig4}
\end{figure}

\begin{figure}[p] 
\begin{center}
\epsfig{file=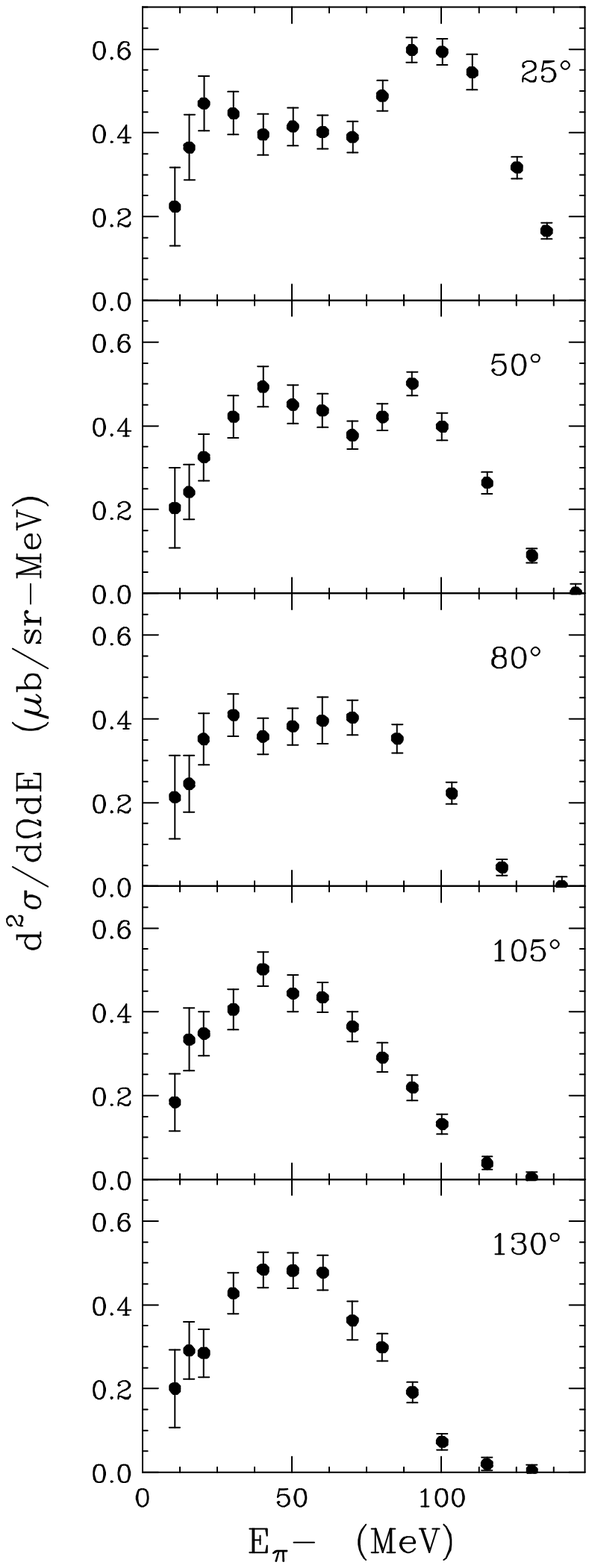,width=0.45\linewidth}
\end{center}
\caption{Doubly differential cross sections for $^4$He$(\pi^+ ,\, \pi^- )$
at 180 MeV for laboratory angles 25$\protect^\circ$, 50$^\circ$,
80$^\circ$, 105$^\circ$, and 130$^\circ$. 
The uncertainties shown
include the statistical uncertainty and the systematic uncertainties which
depend on the outgoing pion energy.}
\label{fig5}
\end{figure}

\begin{figure}[p] 
\begin{center}
\epsfig{file=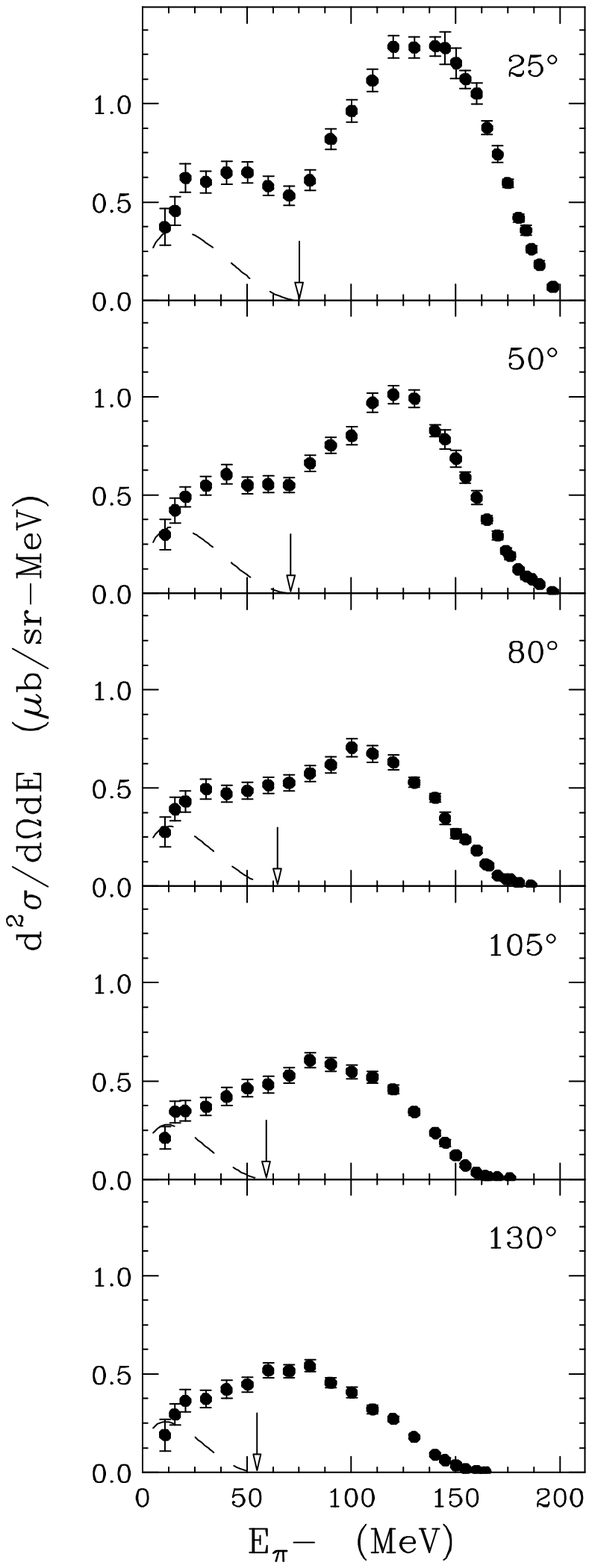,width=0.42\linewidth}
\end{center}
\caption{Doubly differential cross sections for $^4$He$(\pi^+ ,\, \pi^- )$
at 240 MeV for laboratory angles 25$\protect^\circ$, 50$^\circ$,
80$^\circ$, 105$^\circ$, and 130$^\circ$. 
The uncertainties shown
include the statistical uncertainty and the systematic uncertainties which
depend on the outgoing pion energy.  The arrow indicates the upper limit of 
outgoing pion energy from the pion-induced pion production process.  The dashed 
curve represents the shape of the available phase space for pion production.} 
\label{fig6} 
\end{figure}

\begin{figure}[p] 
\begin{center}
\epsfig{file=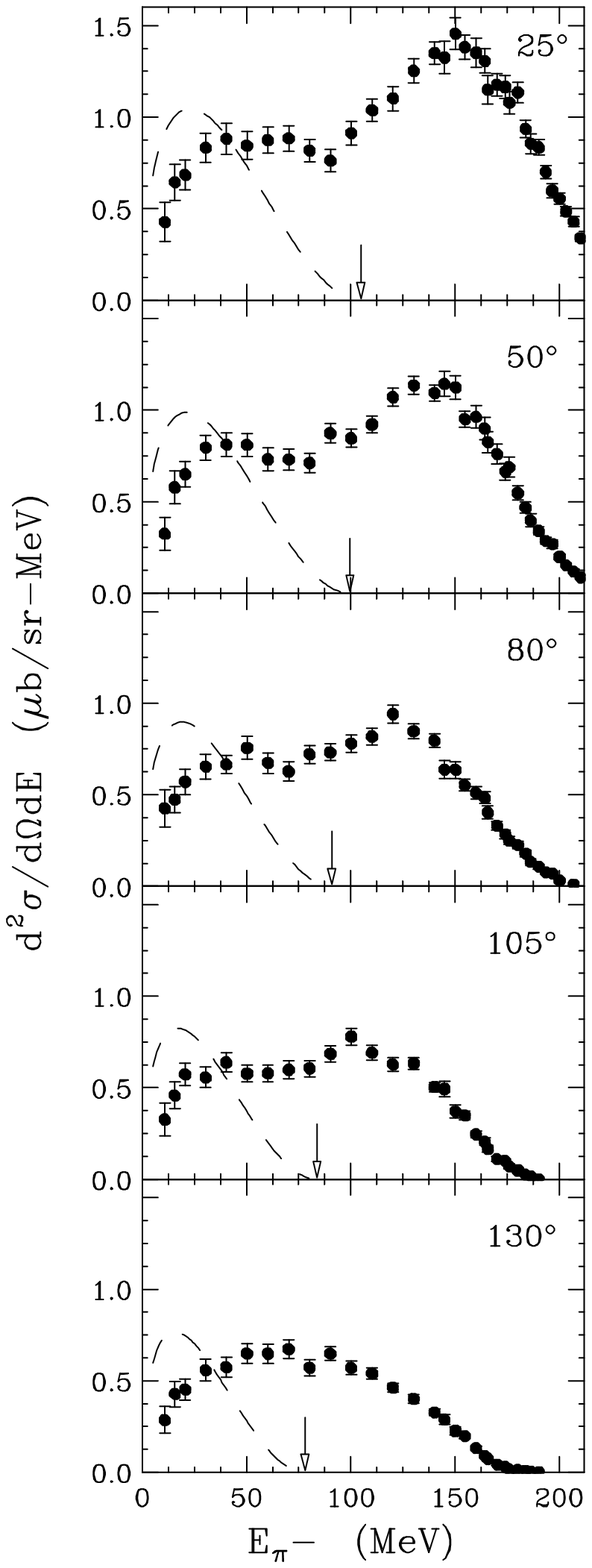,width=0.42\linewidth}
\end{center}
\caption{Doubly differential cross sections for $^4$He$(\pi^+ ,\, \pi^- )$
at 270 MeV for laboratory angles 25$\protect^\circ$, 50$^\circ$,
80$^\circ$, 105$^\circ$, and 130$^\circ$. 
The uncertainties shown
include the statistical uncertainty and the systematic uncertainties which
depend on the outgoing pion energy.  The arrow indicates the upper limit of 
outgoing pion energy from the pion-induced pion production process. The dashed curve 
represents the shape of the available phase space for pion production.}
\label{fig7}
\end{figure}

\begin{figure}[p] 
\begin{center}
\epsfig{file=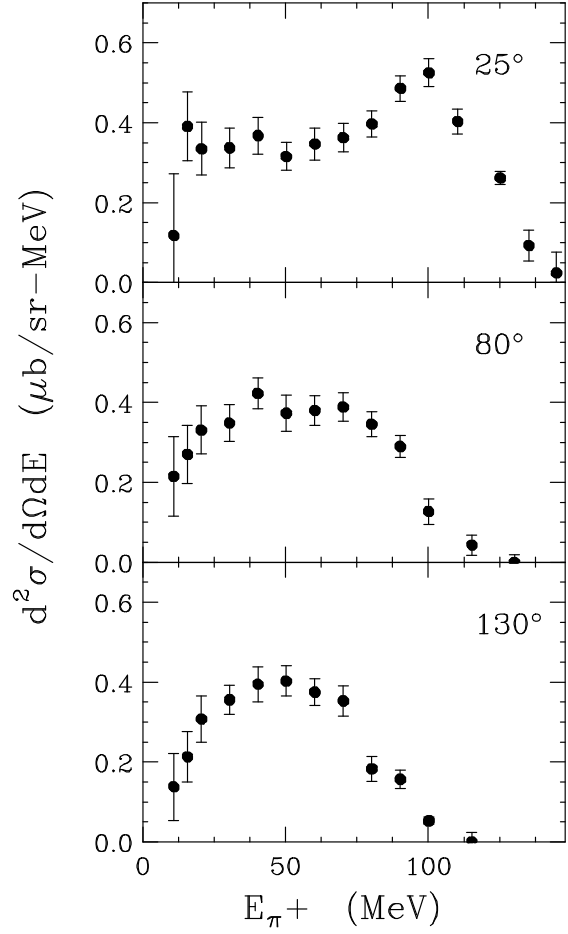,width=0.45\linewidth}
\end{center}
\caption{Doubly differential cross sections for $^4$He$(\pi^-, \pi^+)$ at 
incident pion energy 180 MeV for laboratory angles 25$^\circ$, 
80$^\circ$, and 130$^\circ$.
The uncertainties shown
include the statistical uncertainty and the systematic uncertainties which
depend on the outgoing pion energy.}
\label{fig8}
\end{figure}

\begin{figure}[p] 
\begin{center}
\epsfig{file=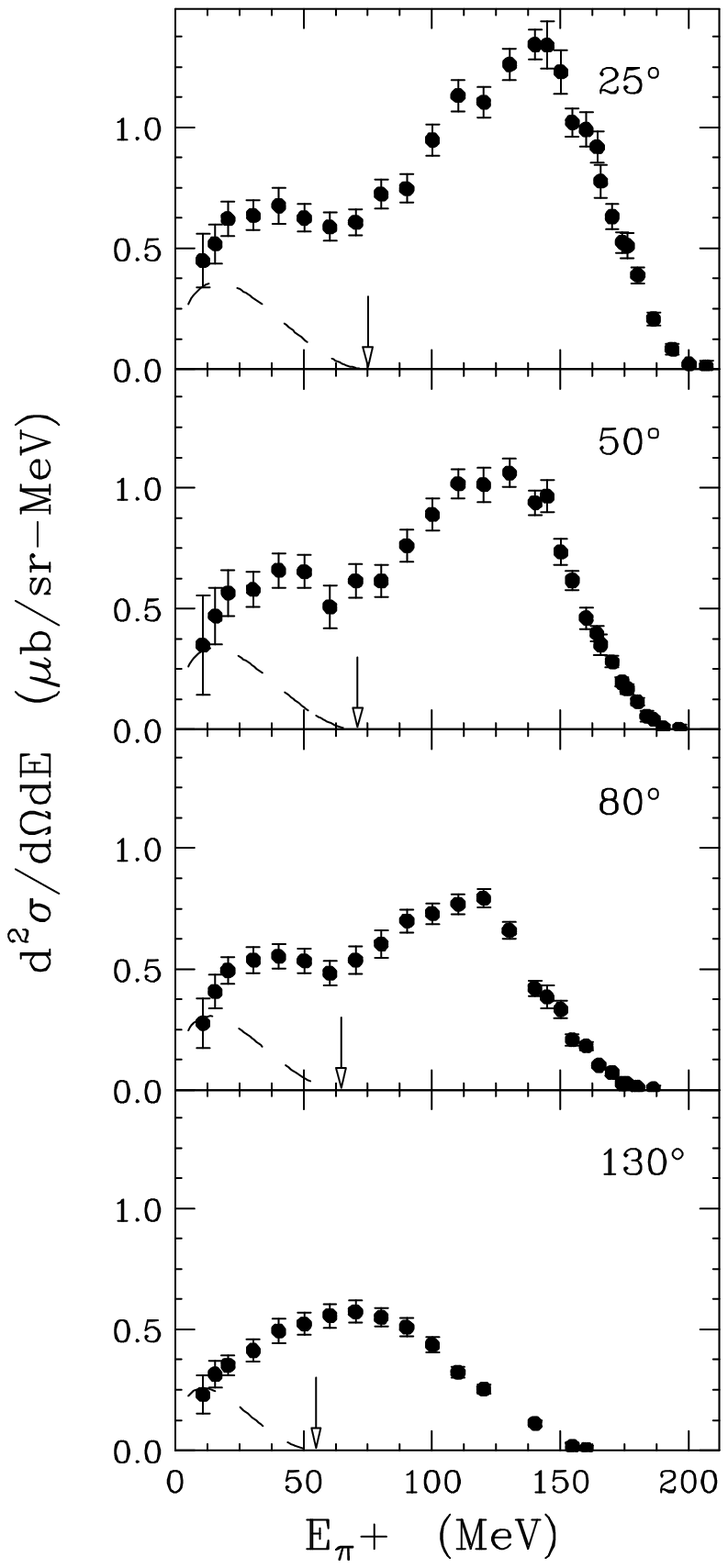,width=0.45\linewidth}
\end{center}
\caption{Doubly differential cross sections for $^4$He$(\pi^-, \pi^+)$ at 
incident pion energy 240 MeV for laboratory angles 25$^\circ$, 50$^\circ$, 
80$^\circ$, and 130$^\circ$.
The uncertainties shown
include the statistical uncertainty and the systematic uncertainties which
depend on the outgoing pion energy. The arrow indicates the upper limit of 
outgoing pion energy from the pion-induced pion production process. The dashed curve 
represents the shape of the available phase space for pion production.}
\label{fig9}
\end{figure}

\begin{figure}[p] 
\begin{center}
\epsfig{file=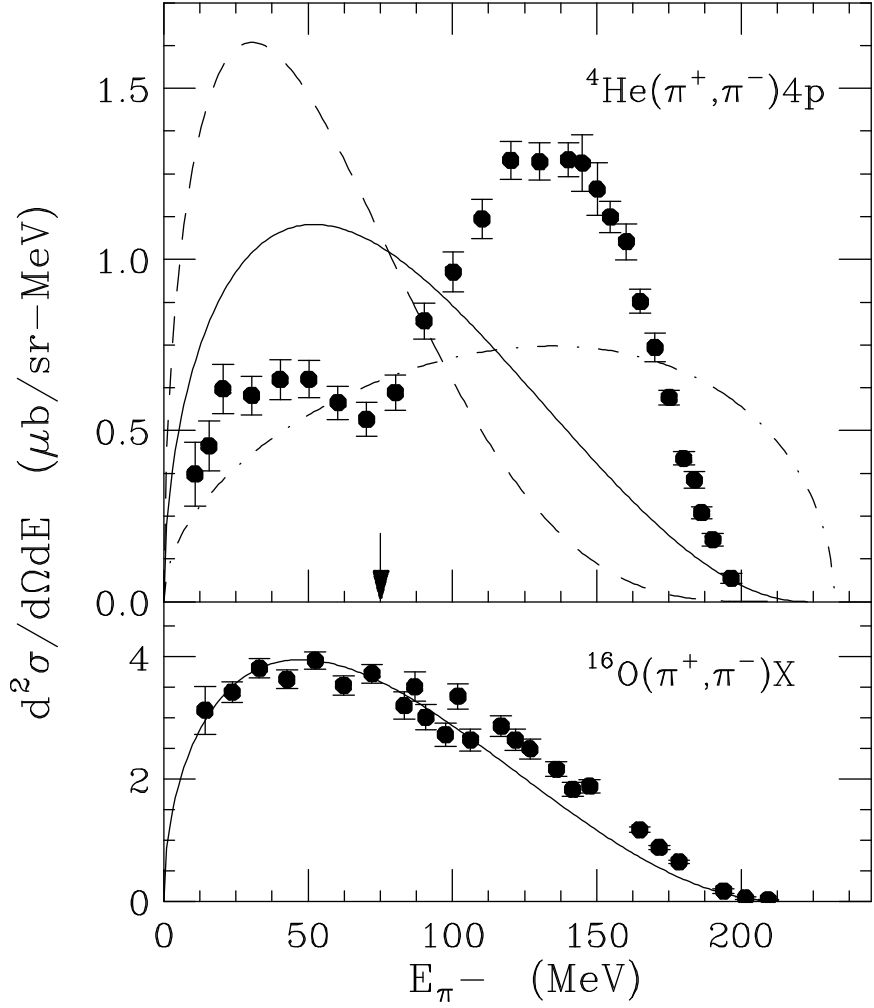,width=0.7\linewidth}
\end{center}
\caption{A comparison of the doubly differential cross sections at 240 MeV 
and 25$^\circ$ for the $(\pi^+, \pi^-)$ reactions in $^4$He and 
$^{16}$O.  The $^{16}$O data are taken from 
Ref.~\protect\cite{wood}.  The 
dashed and dot-dashed curves correspond to the 
distribution of events in five-body and three-body phase space, 
respectively, while the solid curves correspond to 
four-body phase space.  The arrow indicates the maximum pion 
energy allowed in PIPP. The phase space curves are
normalized as described in the text.}
\label{fig10}
\end{figure}

\begin{figure}[p] 
\begin{center}
\epsfig{file=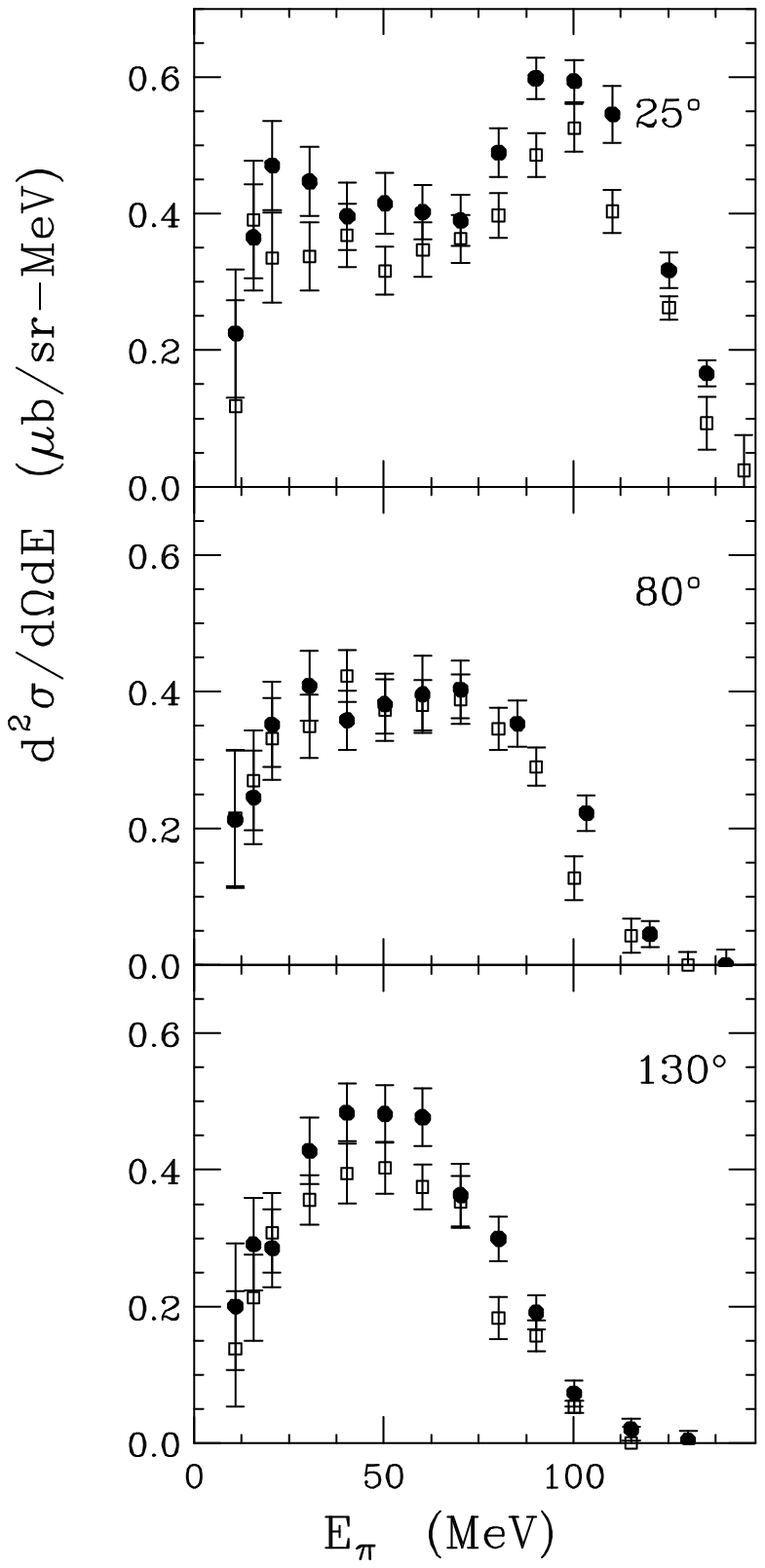,width=0.5\linewidth}
\end{center}
\caption{A comparison of the doubly differential cross sections for 
$^4$He$
(\pi^+, \, \pi^- )$ (solid circles) and $^4$He$(\pi^-, \, \pi^+ )$ 
(open squares) at 180 MeV.}
\label{fig11}
\end{figure}

\begin{figure}[p] 
\begin{center}
\epsfig{file=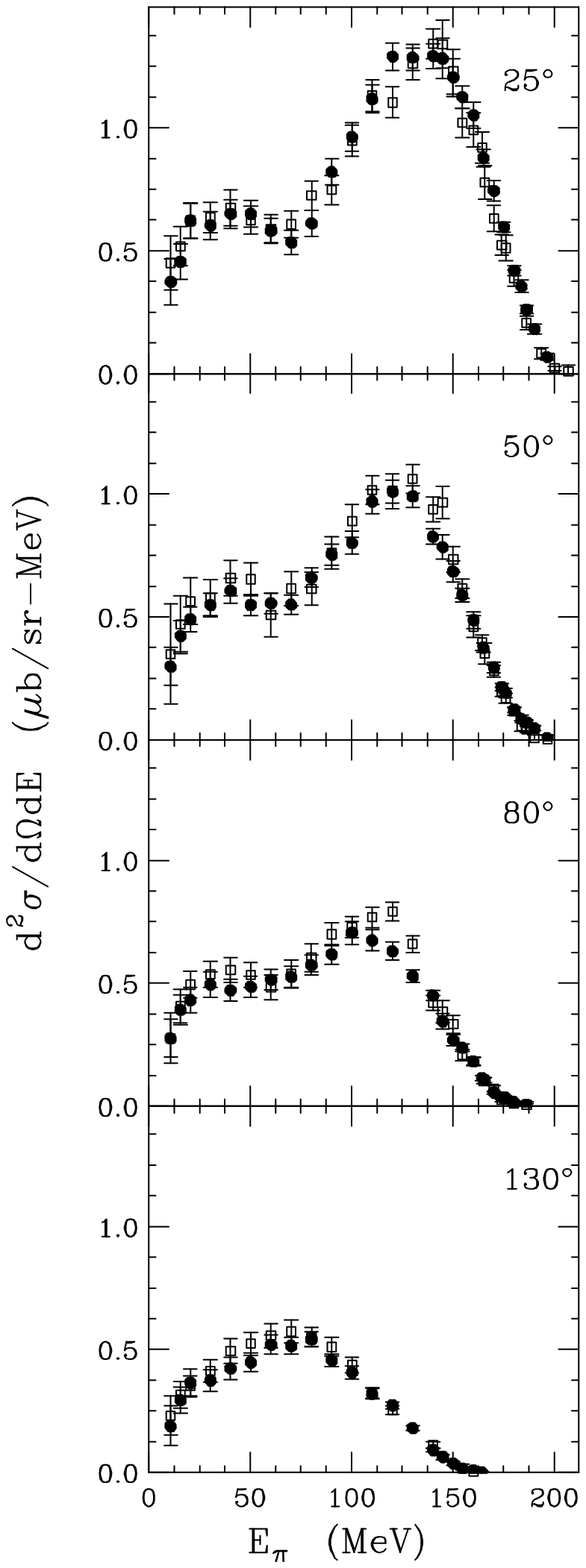,width=0.45\linewidth}
\end{center}
\caption{A comparison of the doubly differential cross sections for 
$^4$He$
(\pi^+, \, \pi^- )$ (solid circles) and $^4$He$(\pi^-, \, \pi^+ )$ 
(open squares) at 240 MeV.}
\label{fig12}
\end{figure}

\begin{figure}[p] 
\begin{center}
\epsfig{file=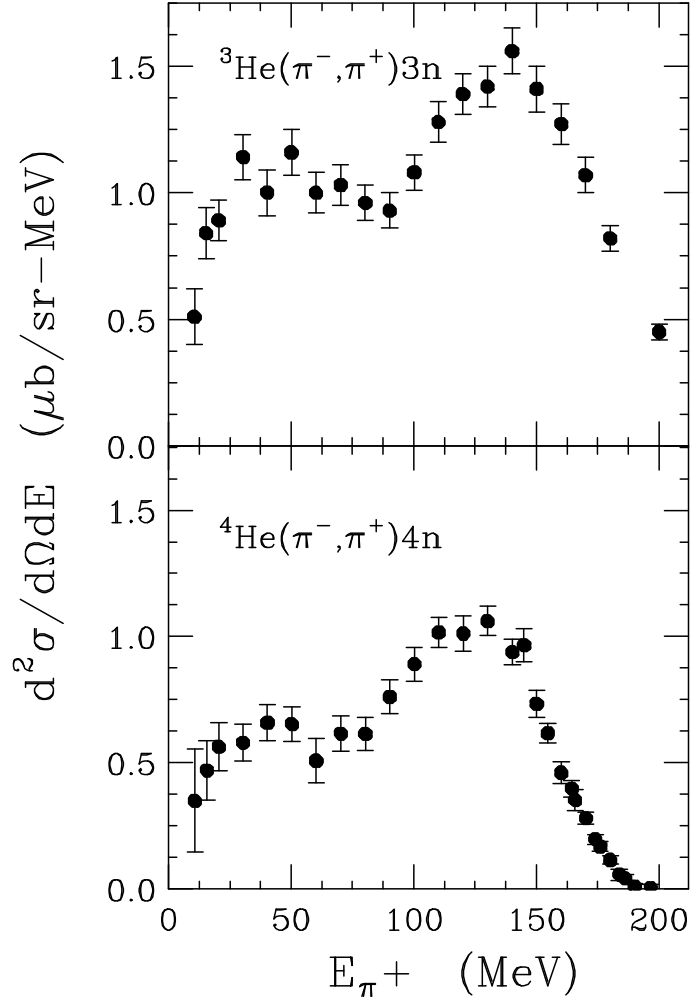,width=0.55\linewidth}
\end{center}
\caption{A comparison of the doubly differential cross sections for 
$^3$He$(\pi^-, \, \pi^+ )$ and $^4$He$(\pi^-, \, \pi^+ )$ 
at 240 MeV and 50$^\circ$.}
\label{fig13}
\end{figure}

\begin{figure}[p] 
\begin{center}
\epsfig{file=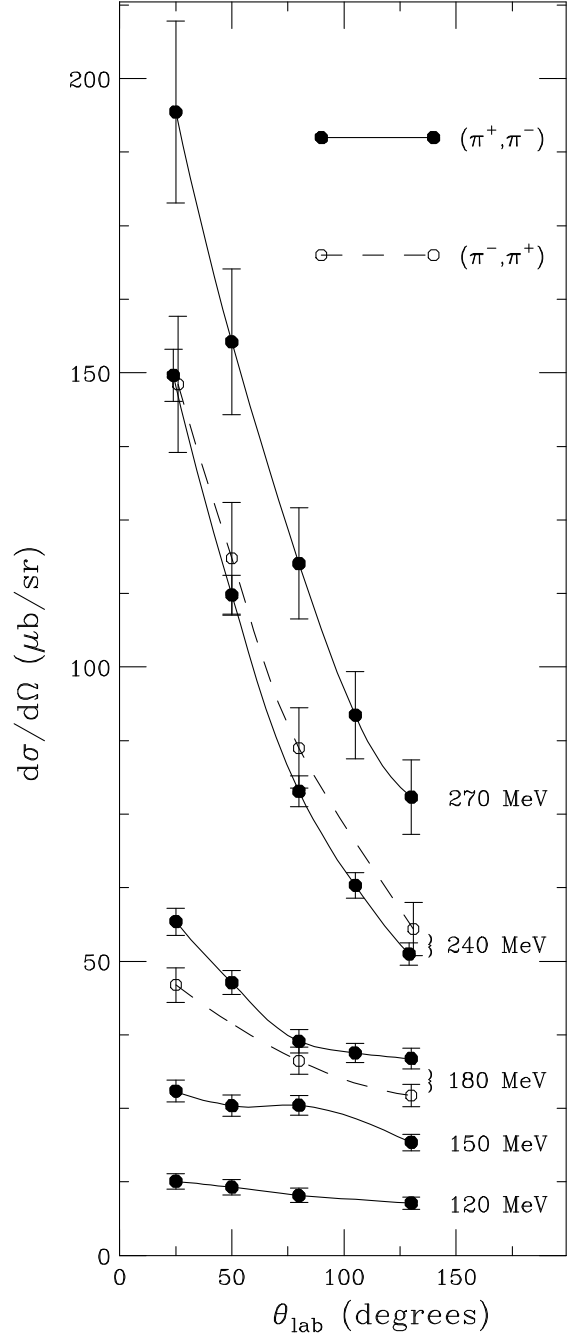,width=0.45\linewidth}
\end{center}
\caption{Angular distributions for $^4$He$(\pi^+, \pi^-)$ (solid circles) 
and $^4$He$(\pi^-, \, \pi^+)$ (open circles)
at 120, 150, 180, 240, and 270 MeV. The uncertainties indicated include 
the statistical uncertainty, the uncertainties arising from the 
extrapolation and integration
procedure (see Sect.~V.G), and
the systematic uncertainties which
depend on the outgoing pion energy and angle.  The curves serve only 
to guide the eye.}
\label{fig14}
\end{figure}

\begin{figure}[p] 
\begin{center}
\epsfig{file=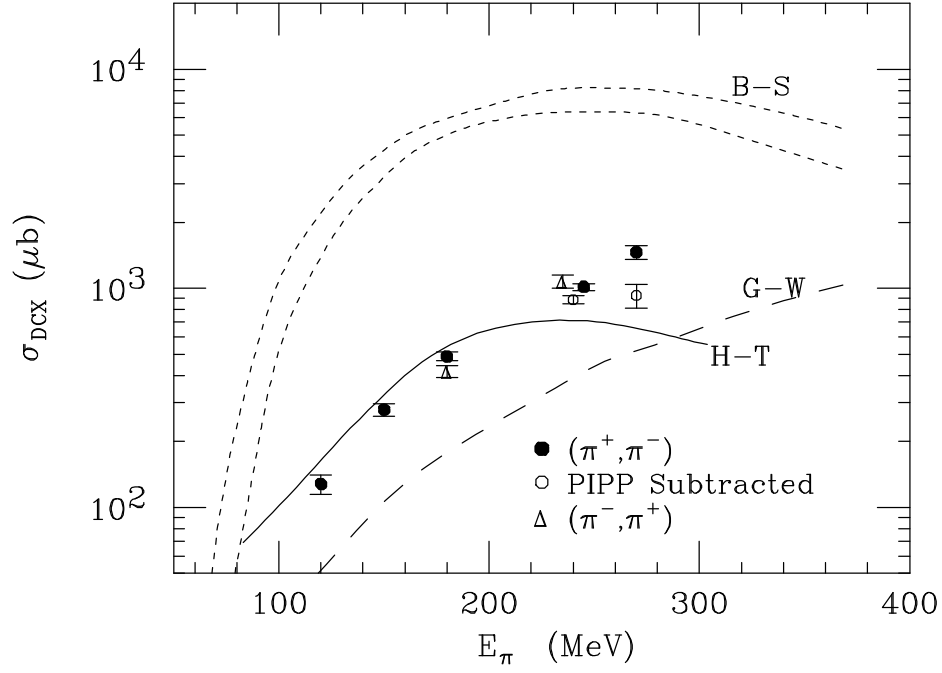,width=0.75\linewidth}
\end{center}
\caption{Total reaction cross sections for $^4$He$(\pi^+, \pi-)$ (solid 
circles) and $^4$He$(\pi^-, \pi^+)$ (open triangles) as a function of 
incident pion energy.  The open circles are the $(\pi^+, \pi^-)$ 
cross sections at 240 MeV and 270 MeV after subtraction of the estimated 
contribution of pion-induced pion production.  The curves represent 
theoretical predictions which are discussed in the text.}
\label{fig15}
\end{figure}

\begin{figure}[p] 
\begin{center}
\epsfig{file=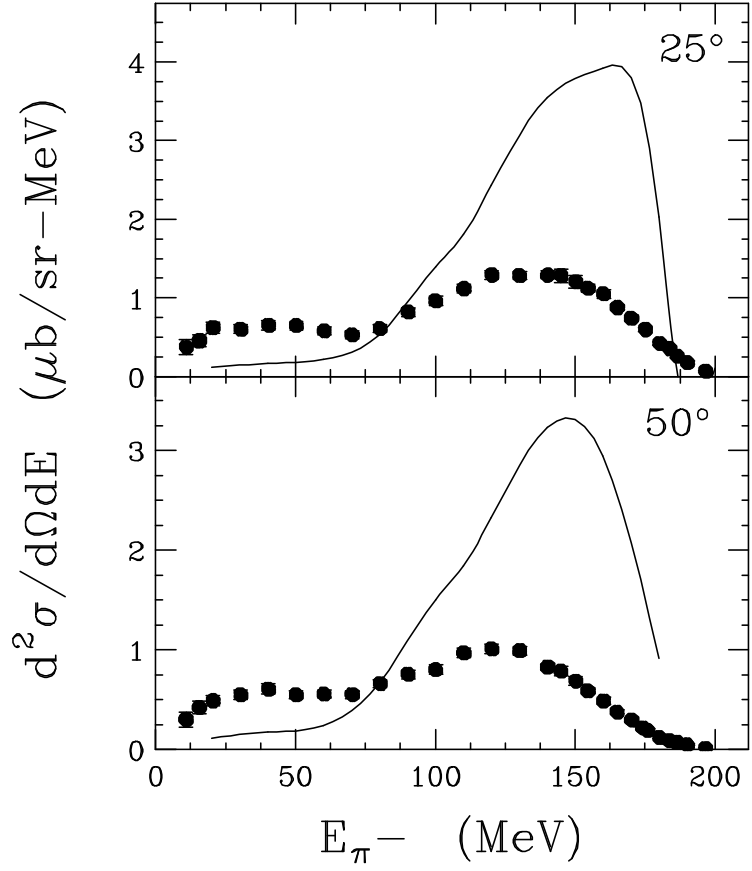,width=0.6\linewidth}
\end{center}
\caption{Comparison of the calculation of Rebka and 
Kulkarni\protect\cite{rebka,kul}, 
based on the 
fixed-nucleon calculation of Gibbs {\it et al.}\protect\cite{gibbs}, with 
the 
$^4$He$(\pi^+, 
\pi^-)$ cross sections at incident energy 240 MeV and laboratory angles 
25$^\circ$ and 50$^\circ$.}
\label{fig16}
\end{figure}

\begin{figure}[p] 
\begin{center}
\epsfig{file=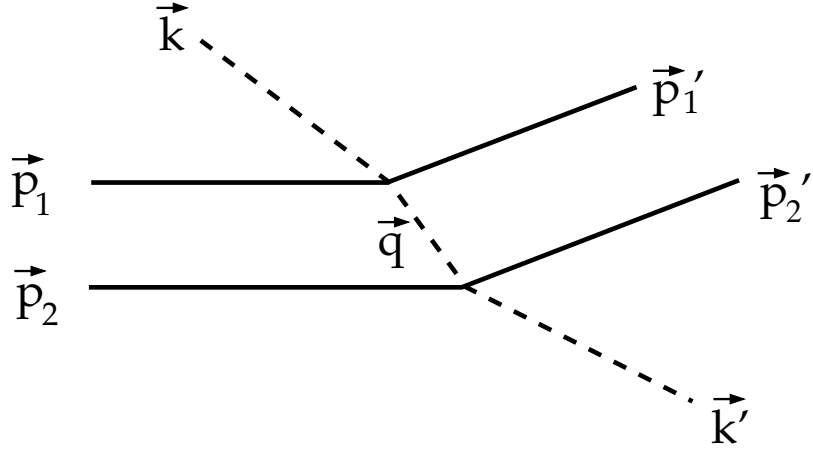,width=0.65\linewidth}
\end{center}
\caption{Diagram of the SSCX process showing the intermediate pion momentum
$\vec q$ and the incident and knocked-out nucleon momenta $\vec p_1$, $\vec
p _2$, $\vec p_1 ^{\; \prime}$, and $\vec p_2 ^{\; \prime}$.}
\label{fig17}
\end{figure}

\clearpage

\begin{figure}[p] 
\begin{center} 
\epsfig{file=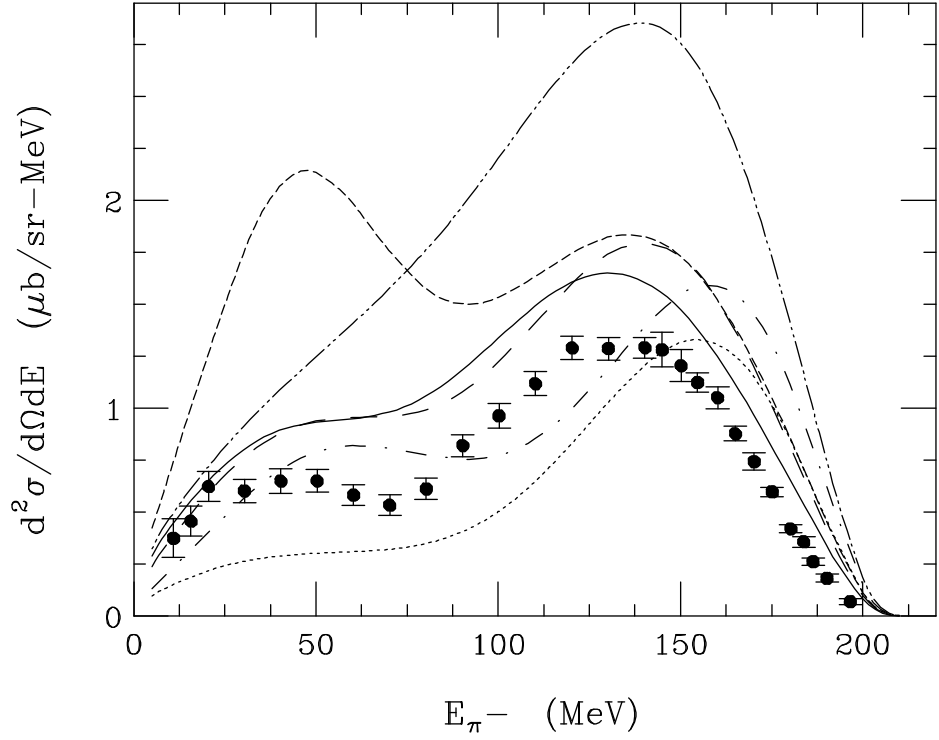,width=0.75\linewidth}
\end{center} 
\caption{Doubly differential cross section for $^4$He$(\pi^+,
\pi^-)$ at 240 MeV and 25$^\circ$ in comparison with predictions of the
non-static sequential single charge exchange model.  The curves represent
different assumptions for the nuclear potentials in the first and second
scattering (see text):  $U_1 = U_2 = -55$ MeV (solid curve); $U_1 = U_2 = -37$
MeV (long-dashed curve); $U_1 = U_2 = 0$ (dot-dashed curve); $U_1 = -55$ MeV, 
$U_2 = 0$ (dotted curve); $U_1 = 0$, $U_2 = -55$ MeV (short-dashed curve).  The
double-dot-dashed curve represents the static prescription (nucleons at rest)
with $U_1 = U_2 = 0$.} 
\label{fig18} 
\end{figure}

\begin{figure}[p] 
\begin{center}
\epsfig{file=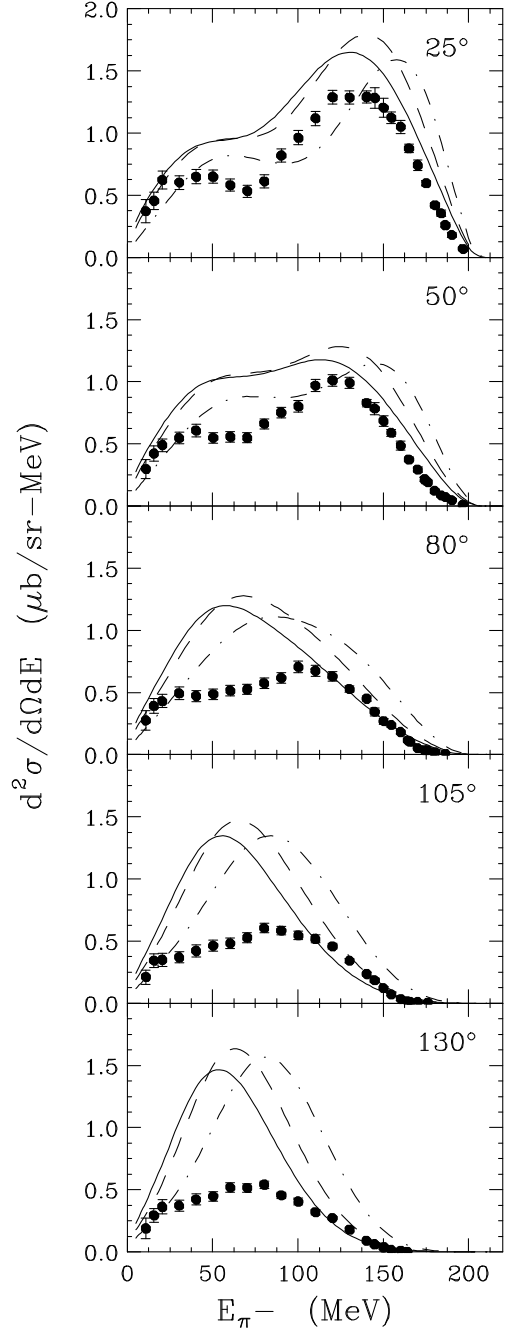,width=0.40\linewidth}
\end{center}
\caption{Doubly differential cross sections for $^4$He$(\pi^+,\pi^-)$ at
240 MeV and laboratory angles 25$^\circ$, 50$^\circ$, 80$^\circ$,
105$^\circ$, and 130$^\circ$.  The theoretical curves are the same as in
Fig.~\protect\ref{fig18} using the average nuclear potentials $U_1 = U_2 =
-55$ MeV (solid curve), $U_1 = U_2 = -37$ MeV (dashed curve), and $U_1 = 
U_2 = 0$ (dot-dashed curve).}
\label{fig19}
\end{figure}

\begin{figure}[p] 
\begin{center}
\epsfig{file=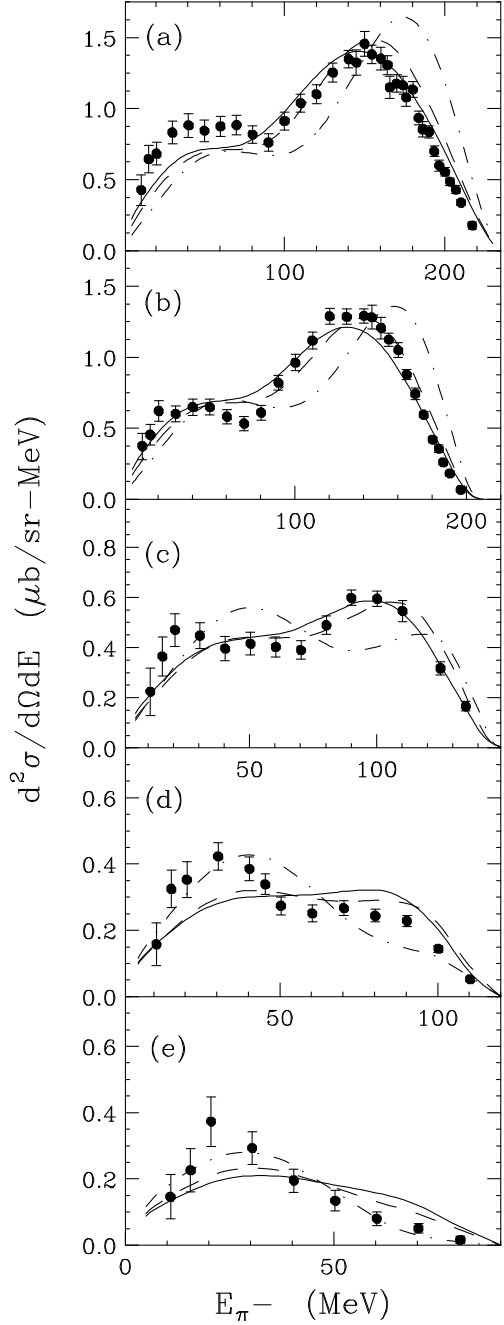,width=0.4\linewidth}
\end{center}
\caption{Doubly differential cross sections for $^4$He$(\pi^+, \pi^-)$ at
laboratory angle 25$^\circ$ for incident energies 270 (a), 240 (b), 180 (c), 
150 (d), and
120 (e) MeV, using the average nuclear potentials $U_1=U_2=-55$ MeV (solid
curve), $U_1=U_2=-37$ MeV (dashed curve), and $U_1=U_2= 0$ (dot-dashed 
curve).  The theoretical predictions have been normalized to yield the 
same
integrated areas as those of the measured cross sections.}
\label{fig20}
\end{figure}

\begin{figure}[p] 
\begin{center}
\epsfig{file=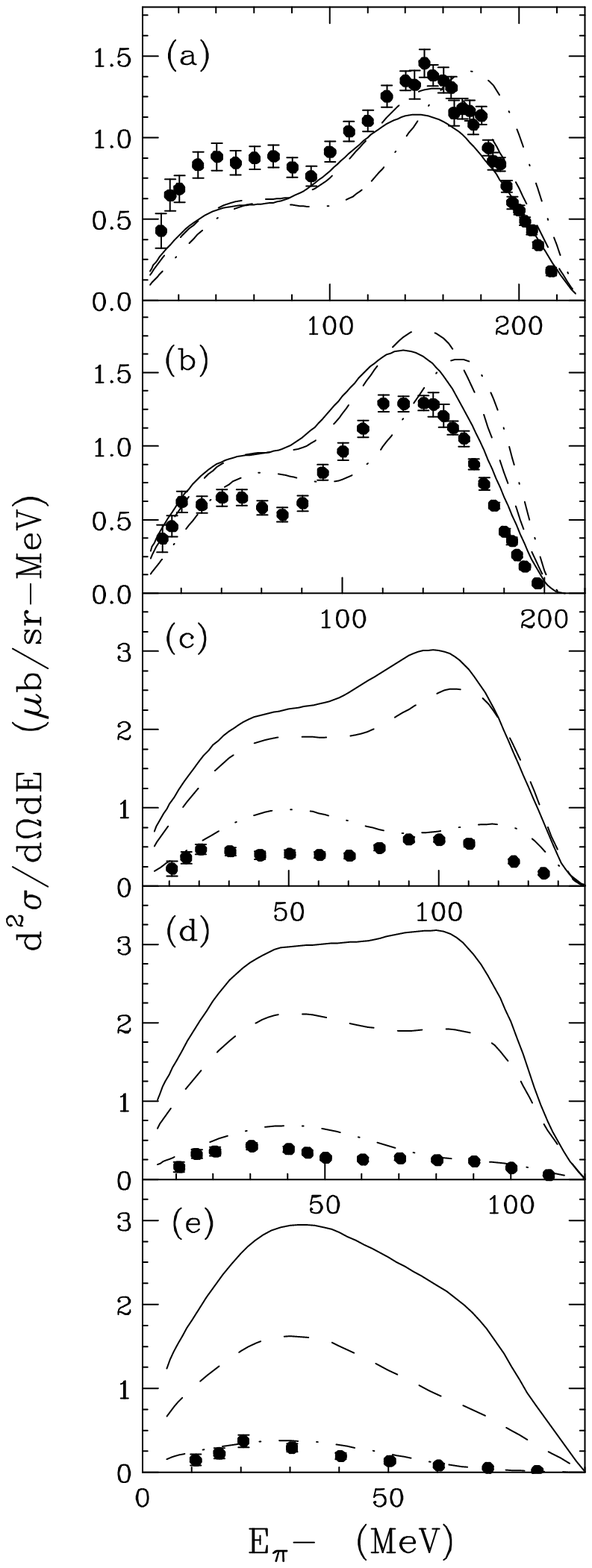,width=0.45\linewidth}
\end{center}
\caption{Doubly differential cross sections for $^4$He$(\pi^+, \pi^-)$ at
laboratory angle 25$^\circ$ for incident energies 270 (a), 240 (b), 180 (c), 
150 (d), and
120 (e) MeV, using the average nuclear potentials $U_1=U_2=-55$ MeV (solid
curve), $U_1=U_2=-37$ MeV (dashed curve), and $U_1=U_2= 0$ (dot-dashed 
curve).  The theoretical predictions have not been normalized.}
\label{fig21}
\end{figure}

\begin{figure}[p] 
\begin{center}
\epsfig{file=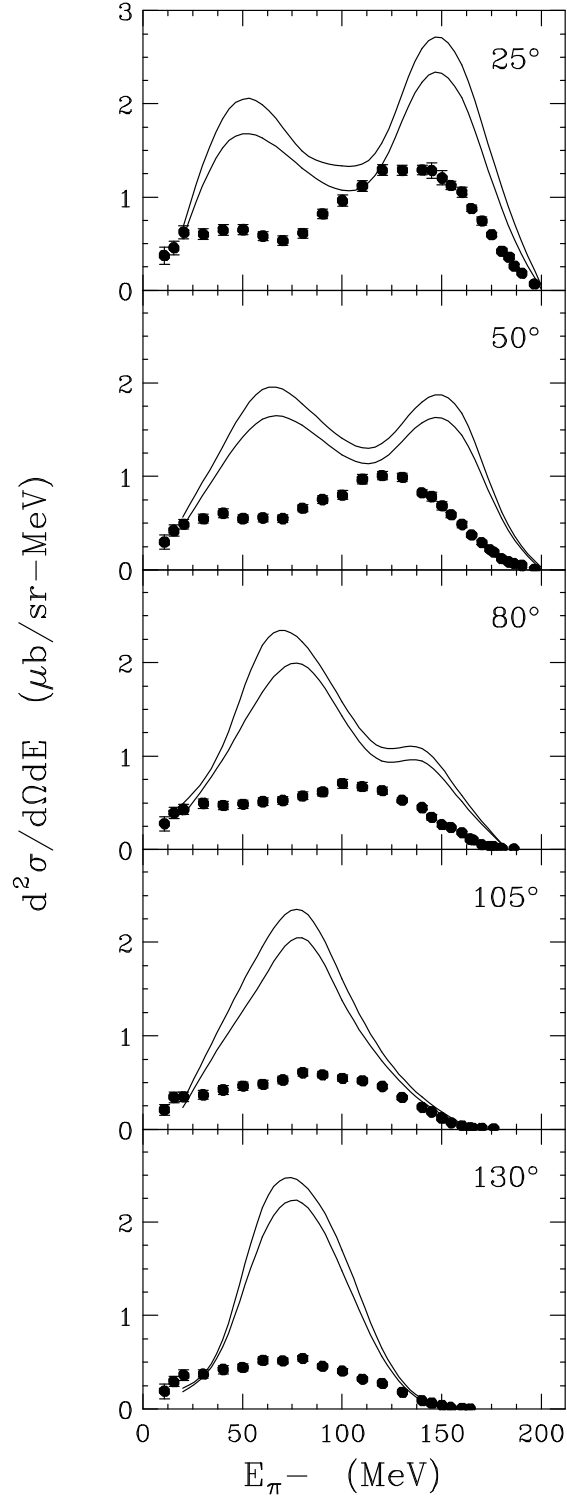,width=0.45\linewidth}
\end{center}
\caption{Comparison of the relativistic three-body model calculation of 
Kulkarni\protect\cite{kul} with the doubly differential cross 
sections for $^4$He$(\pi^+, \pi^-)$ at 240 MeV and laboratory angles 
25$^\circ$, 50$^\circ$, 80$^\circ$, 105$^\circ$, and 130$^\circ$. The upper
and lower curves indicate the uncertainty in the calculation.}
\label{fig22}
\end{figure}

\end{document}